\documentclass[preprint, superscriptaddress, prb]{revtex4-2}
\usepackage{lineno}
\usepackage{graphicx}
\usepackage{float}
\usepackage{amssymb}
\usepackage{mathrsfs}
\usepackage{hyperref}
\usepackage[british]{babel}

\begin{document}

\title{Deterministic multi-qubit entanglement in a quantum network}
\author{Youpeng Zhong}
\thanks{Present Address: Shenzhen Institute for Quantum Science and Engineering, Southern University of Science and Technology, Shenzhen 518055, China}
\affiliation{Pritzker School of Molecular Engineering, University of Chicago, Chicago IL 60637, USA}
\author{Hung-Shen Chang}
\affiliation{Pritzker School of Molecular Engineering, University of Chicago, Chicago IL 60637, USA}
\author{Audrey Bienfait}
\thanks{Present Address: Universit\'{e} de Lyon, ENS de Lyon, Universit\'{e} Claude Bernard, CNRS, Laboratoire de Physique, F-69342 Lyon, France}
\affiliation{Pritzker School of Molecular Engineering, University of Chicago, Chicago IL 60637, USA}
\author{\'Etienne Dumur}
\thanks{Present Address: Universit\'{e} Grenoble Alpes, CEA, INAC-Pheliqs, 38000 Grenoble, France}
\affiliation{Pritzker School of Molecular Engineering, University of Chicago, Chicago IL 60637, USA}
\affiliation{Center for Molecular Engineering and Material Science Division, Argonne National Laboratory, Argonne IL 60439, USA}
\author{Ming-Han Chou}
\affiliation{Pritzker School of Molecular Engineering, University of Chicago, Chicago IL 60637, USA}
\affiliation{Department of Physics, University of Chicago, Chicago IL 60637, USA}
\author{Christopher R. Conner}
\affiliation{Pritzker School of Molecular Engineering, University of Chicago, Chicago IL 60637, USA}
\author{Joel Grebel}
\affiliation{Pritzker School of Molecular Engineering, University of Chicago, Chicago IL 60637, USA}
\author{Rhys G. Povey}
\affiliation{Pritzker School of Molecular Engineering, University of Chicago, Chicago IL 60637, USA}
\affiliation{Department of Physics, University of Chicago, Chicago IL 60637, USA}
\author{Haoxiong Yan}
\affiliation{Pritzker School of Molecular Engineering, University of Chicago, Chicago IL 60637, USA}
\author{David I. Schuster}
\affiliation{Department of Physics, University of Chicago, Chicago IL 60637, USA}
\affiliation{Pritzker School of Molecular Engineering, University of Chicago, Chicago IL 60637, USA}
\author{Andrew N. Cleland}
\email{Corresponding author; anc@uchicago.edu}
\affiliation{Pritzker School of Molecular Engineering, University of Chicago, Chicago IL 60637, USA}
\affiliation{Center for Molecular Engineering and Material Science Division, Argonne National Laboratory, Argonne IL 60439, USA}

\maketitle

\textbf{
Quantum entanglement is a key resource for quantum computation and quantum communication~\cite{Nielsen2010}. Scaling to large quantum communication or computation networks further requires the deterministic generation of multi-qubit entanglement~\cite{Gottesman1999,Duan2001,Jiang2007}. The deterministic entanglement of two remote qubits has recently been demonstrated with microwave photons~\cite{Kurpiers2018,Axline2018,Campagne2018,Leung2019,Zhong2019}, optical photons~\cite{Humphreys2018} and surface acoustic wave phonons~\cite{Bienfait2019}. However, the deterministic generation and transmission of multi-qubit entanglement has not been demonstrated, primarily due to limited state transfer fidelities. Here, we report a quantum network comprising two separate superconducting quantum nodes connected by a 1 meter-long superconducting coaxial cable, where each node includes three interconnected qubits. By directly connecting the coaxial cable to one qubit in each node, we can transfer quantum states between the nodes with a process fidelity of $0.911\pm0.008$. Using the high-fidelity communication link, we can prepare a three-qubit Greenberger-Horne-Zeilinger (GHZ) state~\cite{Greenberger1990,Neeley2010,Dicarlo2010} in one node and deterministically transfer this state to the other node, with a transferred state fidelity of $0.656\pm 0.014$. We further use this system to deterministically generate a two-node, six-qubit GHZ state, globally distributed within the network, with a state fidelity of $0.722\pm0.021$. The GHZ state fidelities are clearly above the threshold of $1/2$ for genuine multipartite entanglement~\cite{Guhne2010}, and show that this architecture can be used to coherently link together multiple superconducting quantum processors, providing a modular approach for building large-scale quantum computers~\cite{Monroe2014,Chou2018}.
}

Superposition and entanglement are key resources enabling both quantum computing and quantum communication. The deterministic generation and distribution of entanglement in a scalable architecture is thus a central requirement underpinning these technologies. Superconducting qubits show great promise as a scalable approach to building practical quantum computers~\cite{Arute2019,Rosenberg2020}, as well as for coherently linking superconducting processors within a cryostat~\cite{Kurpiers2018,Axline2018,Campagne2018,Leung2019} or cryogenically-linked cryostats~\cite{Magnard2020}. Developments in microwave-to-optical transduction promise further extensions of superconducting quantum networks~\cite{Bochmann2013,Mirhosseini2020}, potentially allowing for long-distance quantum communication~\cite{Hensen2015,Liao2017}. However, fundamental challenges still remain. In particular, the fidelity of chip-to-chip quantum state transfers, using microwave-frequency photons, has been limited to $\sim0.8$ due to losses in the communication channels~\cite{Kurpiers2018,Axline2018,Campagne2018,Leung2019,Magnard2020}, although experiments minimizing this loss point to the potential for high-fidelity communication~\cite{Zhong2019,Chang2020,Burkhart2020}. Here, we demonstrate a very low-loss connection between two physically-distant quantum nodes fabricated on separate dies, with which we demonstrate a state transfer fidelity of $0.911\pm0.008$. This allows us to deterministically transfer fully-entangled GHZ states between the two nodes, as well as generate a full two-node entangled state, paving the way for modular approaches to large-scale quantum computing and intra-cryostat quantum communication~\cite{Monroe2014}.

\begin{figure}[t]
\begin{center}
	\includegraphics[width=0.8\textwidth]{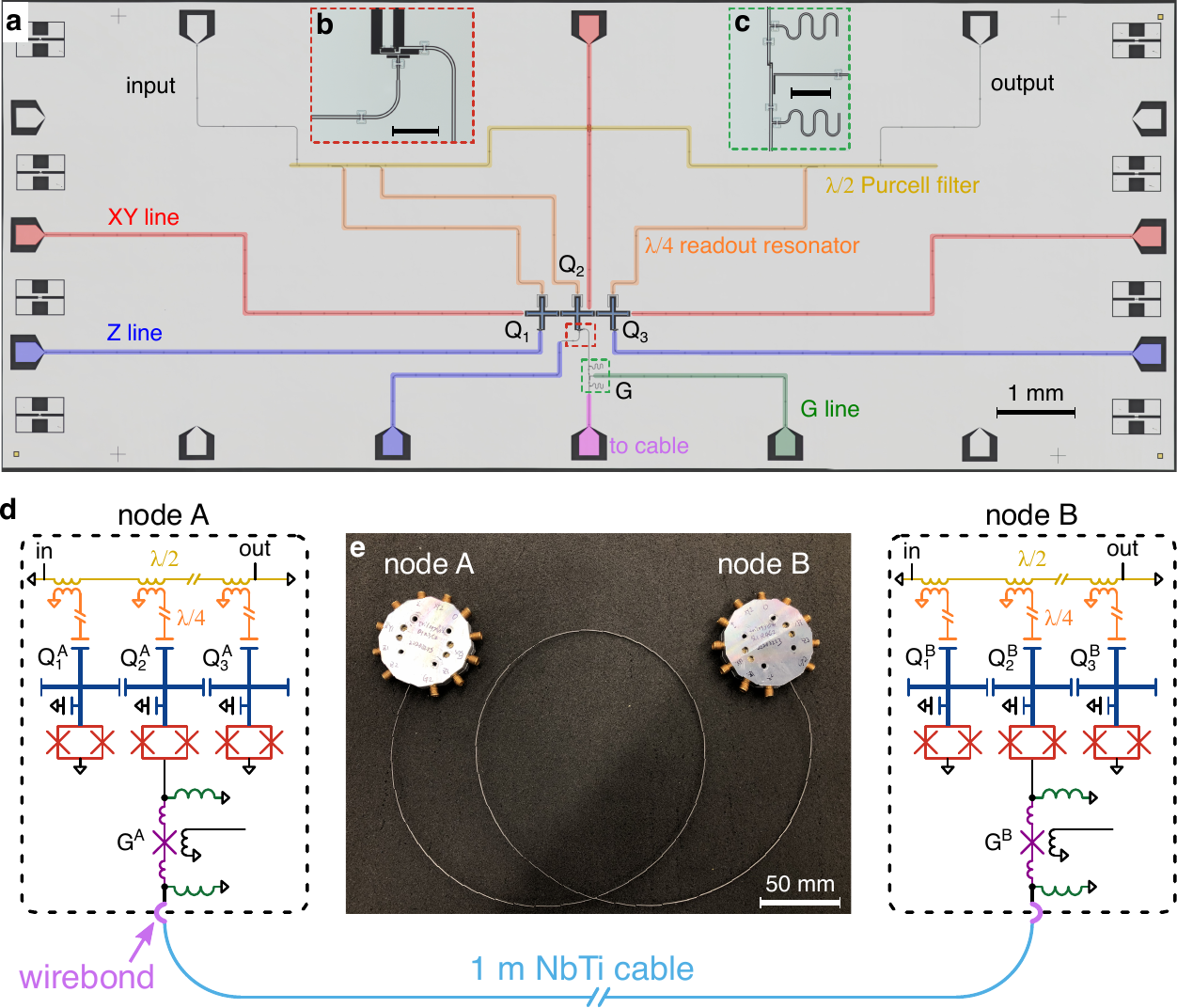}
	\caption{
    \label{fig1}
    {\bf Device description.}
    {\bf a,} False-color micrograph of one quantum processor node, consisting of three capacitively-coupled superconducting qubits $Q_i$ ($i=1,2,3$) with a tunable coupler $G$ connected to $Q_2$. {\bf b, c,} Higher-magnification micrographs of the $Q_2$ Josephson junctions and the tunable coupler $G$, respectively. Scale bars are 100 $\mu$m.
    {\bf d,} Schematic of the quantum network, consisting of two nodes $A$ and $B$ connected by a 1 m-long superconducting NbTi coaxial cable. Each node is a quantum processor of the type shown in panel {\bf a}.
    {\bf e,} Photograph of the quantum network assembly, where each node is in a machined aluminium sample holder connected internally to the superconducting cable.
    }
\end{center}
\end{figure}

Our quantum network consists of two nodes $A$ and $B$, shown in Fig.~\ref{fig1}, where each node is a superconducting processor comprising three capacitively-coupled superconducting qubits $Q_i^n$ ($i=1,2,3$; $n=A,B$), with a tunable coupler~\cite{Chen2014} $G^n$ connected to $Q_2^n$. We use a 1 m-long niobium-titanium (NbTi) superconducting coaxial cable to connect the two nodes together, with a time-variable coupling strength $g^n(t)$ controlled by the tunable coupler $G^n$ in each node. To build a high-quality communication channel, we avoid use of microwave connectors or circulators~\cite{Kurpiers2018,Axline2018,Campagne2018,Leung2019}, relying instead on direct superconducting aluminium wirebond connections between the coaxial cable and the processors; see the Supplementary Information~\cite{SI} for more details.

We place the assembled quantum network in a magnetic shield attached to the mixing chamber of a dilution refrigerator with a base temperature below 10 mK. We first tune up and calibrate the quantum state transfer between $Q_2^A$ and $Q_2^B$, with the other qubits biased far away in frequency. When the coupling is off, the coaxial cable is effectively shorted to ground on both ends, supporting an evenly-spaced sequence of standing microwave modes, with a free spectral range $\omega_{\rm FSR}/2\pi = 105$~MHz. The coupling strength $g^n$ between $Q_2^n$ and each mode is determined~\cite{SI} by the superconducting phase $\delta^n$ across the Josephson junction of the coupler $G^n$. To tune-up each qubit, we isolate the qubits from the cable by biasing the coupler junction to $\delta^n=\pi/2$, turning off the coupling, $g^n \approx 0$. We find each qubit has an intrinsic lifetime of $T_1\approx 10$~$\mu$s and a dephasing time of $T_{\phi}\approx 3$~$\mu$s; see Table S1~\cite{SI} for details.

\begin{figure}[t]
\begin{center}
	\includegraphics[width=0.8\textwidth]{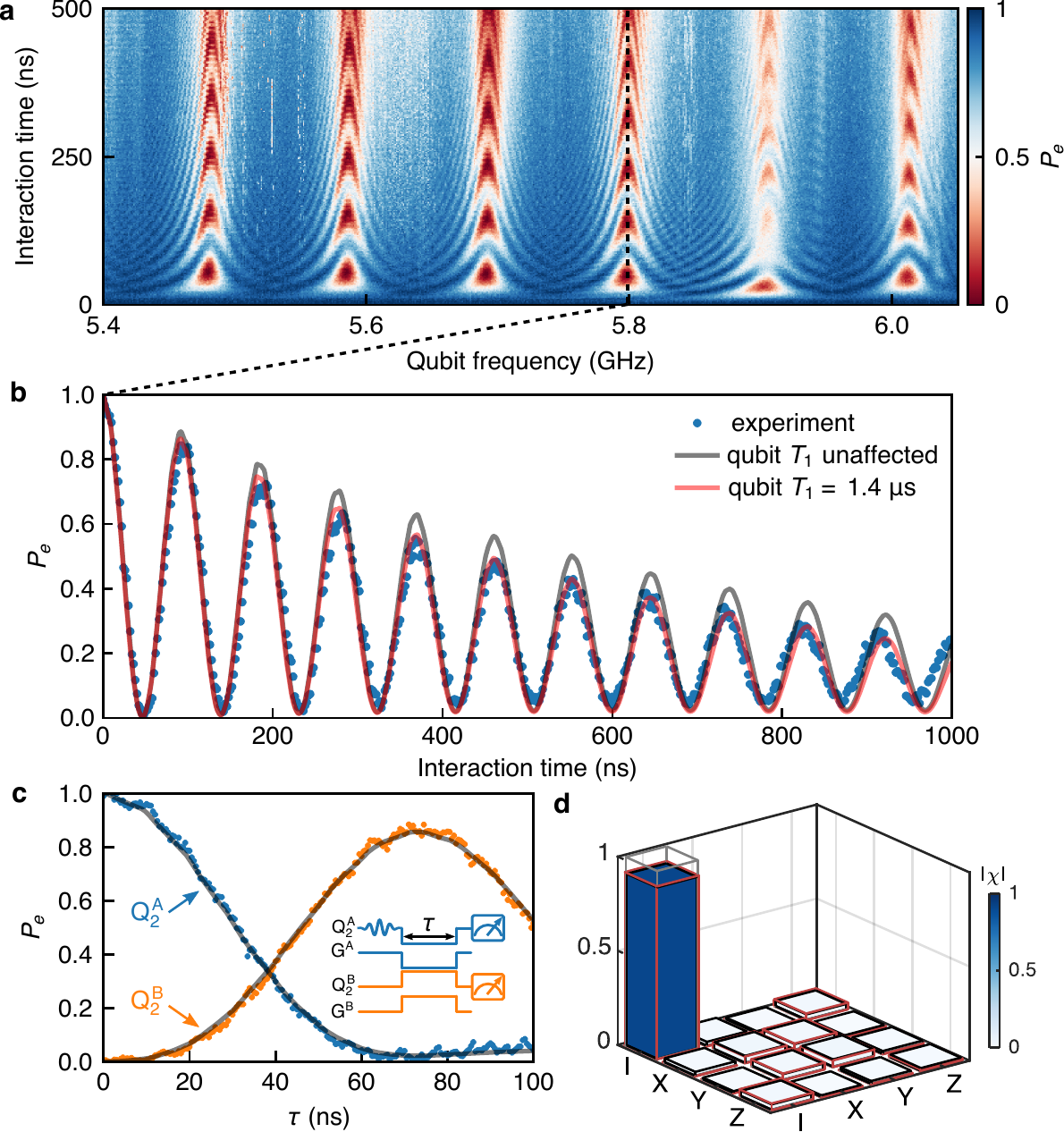}
	\caption{
    \label{fig2}
    {\bf Quantum state transfer between node $A$ and node $B$.}
    {\bf a,} Vacuum Rabi oscillations between $Q_2^A$ and six standing modes in the coaxial cable, with the coupling strength $g^A/2\pi \approx 5.5$~MHz.
    {\bf b,} Slice through data in \textbf{a}, showing the vacuum Rabi oscillation of $Q_2^A$ with the communication mode $R$ at 5.798 GHz. Numerical simulations suggest the effective qubit $T_1$ is shortened to 1.4~$\mu$s during the interaction, due to loss associated with the cable connections~\cite{SI}.
    {\bf c,} Quantum state transfer from node $A$ to node $B$ using a hybrid transfer scheme, where $Q_2^A$ and $Q_2^B$ are resonantly coupled to $R$ with the same coupling strength $g_0/2\pi = 4$ MHz for a duration $\tau$. At $\tau=72$~ns, we achieve a single photon transfer efficiency of $\eta=0.881\pm0.008$. Grey lines: Numerical simulations. Inset: Control pulse sequence.
    {\bf d,} Process matrix $\chi$ for the quantum state transfer, corresponding to a process fidelity $\mathcal{F}^p = 0.911\pm0.008$. The solid bars and red and grey frames are the measured, simulated and ideal values respectively.
    }
\end{center}
\end{figure}

When we prepare the qubit $Q_2^A$ in its excited state $|e\rangle$ and subsequently turn on the coupler $G^A$ to $g^A/2\pi\approx 5.5$~MHz, we observe a sequence of vacuum Rabi oscillations between $Q_2^A$ and the cable standing modes as we vary the qubit frequency and the interaction time, shown in Fig.~\ref{fig2}a. As the mode at 5.798 GHz (dashed line) has a slightly longer lifetime ($T_{1r}=473$ ns) than the other modes, we use this as the communication mode $R$. The on-resonant vacuum Rabi oscillation between $Q_2^A$ and $R$ is shown in detail in Fig.~\ref{fig2}b; more details are provided in the Supplementary Information~\cite{SI}.

If both qubits $Q_2^A$ and $Q_2^B$ are resonantly-coupled to $R$, the tripartite system has a ``dark" and two ``bright" eigenmodes, with very little occupation of the cable in the dark eigenmode~\cite{Chang2020}. As proposed in Ref.~\onlinecite{Wang2012} and demonstrated in Refs.~\onlinecite{Leung2019,Chang2020}, high-fidelity quantum state transfers can be achieved using the dark eigenmode even in the presence of significant cable loss, albeit with limited transfer rates. As we have both cable and qubit loss, we implement a hybrid state transfer scheme~\cite{Wang2012}, which involves all three eigenmodes in a way that balances these different losses. The hybrid scheme involves setting both $G^A$ and $G^B$ to the same coupling strength $g_0/2\pi = 4$ MHz while tuning both $Q_2^A$ and $Q_2^B$ to be resonant with $R$ for a duration $\tau$, shown in Fig.~\ref{fig2}c. At $\tau=72$ ns, one photon is transferred from node $A$ to node $B$ with an efficiency $\eta=0.881\pm0.008$; numerical simulations including the measured loss are in excellent agreement with the measurements~\cite{SI}. We perform quantum process tomography to characterize this state transfer (ST) process, yielding the process matrix $\chi$ shown in Fig.~\ref{fig2}d, with a process fidelity $\mathcal{F}^p = \rm{Tr}(\chi \cdot \chi_{\rm ideal})=0.911\pm0.008$, where $\chi_{\rm ideal}$ is the process matrix for the identity operation $\mathcal{I}$. Numerical simulations give a process fidelity of $0.920$. These simulations imply that the ST process could be further improved by reducing loss associated with the cable and its interconnects~\cite{SI}.

\begin{figure}[t]
\begin{center}
	\includegraphics[width=0.8\textwidth]{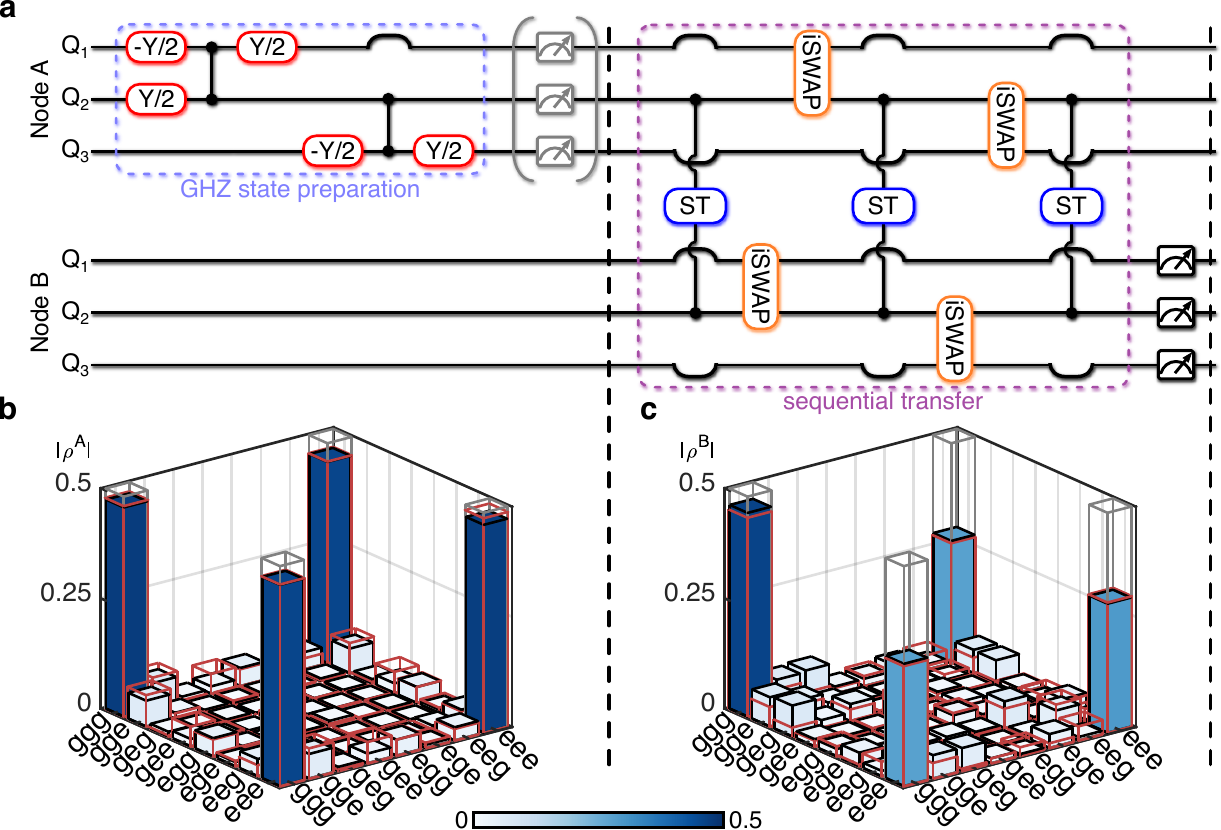}
	\caption{
    \label{fig3}
    {\bf Deterministic transfer of a three-qubit GHZ state.}
    {\bf a,} Schematic of the GHZ state preparation and the sequential transfer protocol. Bumps in the horizontal lines are detuning pulses applied to $Q_j^n$ ($j=1,3$) to minimize interactions between these qubits and $Q_2^n$ during the ST and CZ operations. Measurement of the qubits in node $A$ is only done to characterize the prepared GHZ state, and is not performed when transferring the state to node $B$.
    {\bf b,} Density matrix $\rho^A$ of the GHZ state prepared in node $A$, with a state fidelity of $0.931\pm0.012$.
    {\bf c,} Density matrix $\rho^B$ of the state received in node $B$, with a state fidelity of $0.656\pm 0.014$.
    Solid bars and red and grey frames are measured, simulated, and ideal values, respectively.
    }
\end{center}
\end{figure}

In addition to the single-qubit state transfer process, we tune up controlled-NOT (CNOT) gates built from controlled-Z (CZ) gates combined with single-qubit $\pi/2$ rotations, as well as tuning up iSWAP gates. The iSWAP gate is implemented by biasing $Q_j^n$ ($j=1,3$) to be resonant with $Q_2^n$ for a duration  $\tau_{\rm swap} = \pi/2g^n_{j,2} \approx 15$~ns, where $g^n_{j,2}/2\pi \approx 16.7$~MHz is the capacitive coupling strength between $Q_j^n$ and $Q_2^n$. The CZ gate is implemented~\cite{Strauch2003} by biasing $Q_j^n$ to be resonant with the $|e\rangle$-$|f\rangle$ transition frequency of $Q_2^n$ for a duration $\tau_{\rm CZ} = \pi/\sqrt{2} g^n_{j,2} \approx 21$~ns, completing an $|ee\rangle \rightarrow -i |gf\rangle \rightarrow -|ee\rangle$ process that acquires an overall phase of $\pi$ for the $|ee\rangle$ state, leaving the other basis states unchanged. We characterize~\cite{SI} the CZ gate using quantum process tomography, obtaining a process matrix $\chi_{CZ}$ with an average process fidelity of $0.950\pm 0.006$.

We use these gates to deterministically generate a GHZ state~\cite{Greenberger1990,Neeley2010,Dicarlo2010} in node $A$, $|\psi_{\rm GHZ}\rangle = (|ggg\rangle+|eee\rangle)\sqrt{2}$ (written as $|Q_1^A Q_2^A Q_3^A\rangle$), which we then transfer to node $B$, using the protocol shown in Fig.~\ref{fig3}a. This involves two CNOT gates to prepare the state followed by three sequential transfers (ST) through the cable using $Q_2^n$ ($n=A,B$), interleaved with iSWAPs with $Q_1^n$ or $Q_3^n$.

The density matrix $\rho^A$ of the three-qubit GHZ state in node $A$ is measured using quantum state tomography and shown in Fig.~\ref{fig3}b, with a state fidelity $\mathcal{F}^A = \langle \psi_{\rm GHZ}|\rho^A|\psi_{\rm GHZ}\rangle = 0.931 \pm 0.012$. Calculations using $\chi_{CZ}$ give a state fidelity of $0.938$, in good agreement with the experiment. This state is then transferred to node $B$ using three sequential transfers with interleaved iSWAP gates, yielding the final state $\rho^B$ in node $B$, as shown in Fig.~\ref{fig3}c, with a GHZ state fidelity $\mathcal{F}^B=0.656\pm0.014$, clearly above the threshold of $1/2$ for genuine multipartite entanglement~\cite{Guhne2010}. A calculation applying $\chi^{\otimes 3}$ and the expected decoherence process to $\rho^A$ gives a state fidelity of $0.648$, agreeing well with experiment.

\begin{figure}[t]
\begin{center}
	\includegraphics[width=0.8\textwidth]{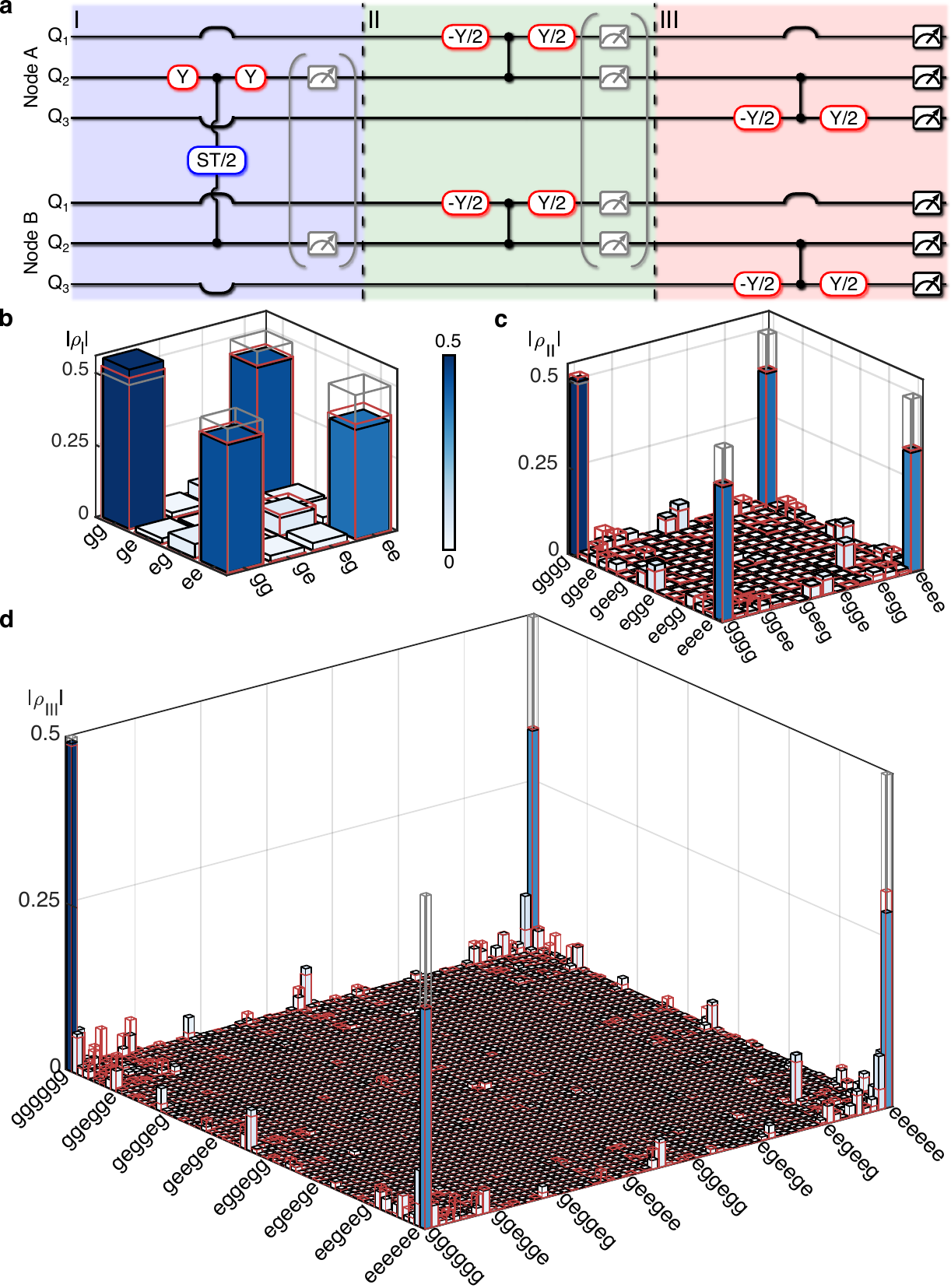}
	\caption{
    \label{fig4}
    {\bf Deterministic generation of multi-qubit entanglement in a quantum network.}
    {\bf a,} The step-by-step protocol for entangling the two nodes $A$ and $B$. Here the ST/2 process involves transmitting half a photon from $Q_2^A$ to $Q_2^B$. Bumps in the horizontal lines are detuning pulses.
    {\bf b,} The Bell triplet state between $Q_2^A$ and $Q_2^B$ created in step I, with a state fidelity of $0.908\pm0.012$.
    {\bf c,} The four-qubit GHZ state created in step II, with a state fidelity of $0.814\pm0.008$.
    {\bf d,} The six-qubit GHZ state created in step III, with a state fidelity of $0.722\pm0.021$.
    The solid bars and red and grey frames are the measured, simulated and ideal values respectively.
    }
\end{center}
\end{figure}

Finally, we demonstrate the step-by-step generation of a six-qubit entangled state distributed in the network, using the protocol shown in Fig.~\ref{fig4}a. In step I, we prepare a Bell triplet state $|B+\rangle=(|gg\rangle+|ee\rangle)/\sqrt{2}$ (written as $|Q_2^A Q_2^B\rangle$), using an ST/2 process similar to the ST process, except the qubit frequencies and coupling parameters are adjusted~\cite{SI} so that an optimal Bell state fidelity is achieved at $\tau=62.8$~ns. The density matrix $\rho_I$ for this process is shown in Fig.~\ref{fig4}b, with a state fidelity of $0.908\pm0.012$. Numerical simulations~\cite{SI} yield a state fidelity of $0.915$. In step II, we apply CNOT gates between $Q_1^n$ and $Q_2^n$ to transform the Bell state into a four-qubit GHZ state $(|gggg\rangle+|eeee\rangle)/\sqrt{2}$ (written as $|Q_1^A Q_2^A Q_1^B Q_2^B\rangle$), with a density matrix $\rho_{II}$ displayed in Fig.~\ref{fig4}c with state fidelity $0.814\pm0.008$. We finally apply CNOT gates between $Q_2^n$ and $Q_3^n$, creating a six-qubit GHZ state $(|gggggg\rangle+|eeeeee\rangle)/\sqrt{2}$ (written as $|Q_1^A Q_2^A Q_3^A Q_1^B Q_2^B Q_3^B\rangle$). The density matrix $\rho_{III}$ of the entangled state is shown in Fig.~\ref{fig4}d, with a state fidelity of $0.722\pm0.021$, clearly above the threshold of $1/2$ for genuine multipartite entanglement~\cite{Guhne2010}. Numerical calculations~\cite{SI} give $\rho_{II}$ and $\rho_{III}$ with state fidelities of $0.829$ and $0.738$ respectively, agreeing well with the experiment.

In conclusion, we have built a two-node quantum network consisting of two three-qubit superconducting processor nodes connected by a 1 m-long superconducting coaxial cable. Using this system, we achieve a state transfer process fidelity of $0.911\pm0.008$ between the two nodes, which supports the deterministic generation and transfer of multi-qubit GHZ states. The transfer fidelity here is primarily limited by loss in the cable connections; improving these connections should yield significant increases in the channel coherence and transfer fidelities. This architecture can be extended to coherently link more than two processor nodes, providing a modular solution for building large-scale quantum computers~\cite{Monroe2014,Chou2018}.

\clearpage

\bibliographystyle{naturemag}
\bibliography{bibliography}

\clearpage

\subsection*{Acknowledgements}
We thank P. J. Duda, A. Clerk and K. J. Satzinger for helpful discussions. We thank W. D. Oliver and G. Calusine at MIT Lincoln Lab for providing the traveling wave parametric amplifier (TWPA) used in this work. Devices and experiments were supported by the Air Force Office of Scientific Research and the Army Research Laboratory. \'E.D. was supported by LDRD funds from Argonne National Laboratory; A.N.C. was supported in part by the DOE, Office of Basic Energy Sciences, and D.I.S. acknowledges support from the David and Lucile Packard Foundation. This work was partially supported by UChicago's MRSEC (NSF award DMR-201185) and made use of the Pritzker Nanofabrication Facility, which receives support from SHyNE, a node of the National Science Foundation's National Nanotechnology Coordinated Infrastructure (NSF Grant No. NNCI1542205).

\subsection*{Author contributions}
Y.P.Z. designed and fabricated the devices, performed the measurement and analyzed the data. H.-S.C. provided help with the measurement. A.B. and \'E.D. provided help with the infrastructure setup. A.N.C. and D.I.S. advised on all efforts. All authors contributed to discussions and production of the manuscript.

\subsection*{Competing interests}
The authors declare no competing interests.

\subsection*{Methods}

\textbf{Quantum state transfer and remote entanglement protocols.}
Probabilistic remote entanglement has been realized with atomic ensembles~\cite{Julsgaard2001,Chou2005}, single atoms~\cite{Moehring2007,Ritter2012}, defects in diamond~\cite{Lee2011,Bernien2013,Hensen2015} and superconducting qubits~\cite{Roch2014,Narla2016,Dickel2018,Kurpiers2019}. However these probabilistic approaches are not scalable; a deterministic approach is preferable for large-scale quantum networks. For short-distance communication, for example with microwave cables shorter than a few meters, the free spectral range of the cable is large enough that a single standing mode can relay quantum states~\cite{Sillanpaa2007,Majer2007,Zhong2019}, or state transfers can be via a ``dark'' mode hybridized by a standing mode and the on-chip elements (qubits or resonators)~\cite{Leung2019,Chang2020}. Here, we use a hybrid scheme~\cite{Wang2012} for state transfer, which balances the loss in the channel with that in the qubits, which might be further improved by optimizing the transfer process using shortcut-to-adiabatic-passage protocols~\cite{Baksic2016,Zhou2017}.

If the length of the cable is increased, the free spectral range of the standing modes in the cable becomes smaller, making single-mode quantum state transfers more challenging. For long-distance communication, the use of itinerant photons is preferable~\cite{Cirac1997,Korotkov2011,Sete2015} but challenging in practice. Using tunable couplers to shape the photon emission and capture in a time-reversal symmetric manner, high-fidelity quantum state transfers have been achieved with itinerant photons~\cite{Yin2013,Wenner2014,Srinivasan2014,Pechal2014,Zeytinouglu2015,Kurpiers2018,Axline2018,Campagne2018}; proposals using a chiral communication channels point to the potential for quantum state transfers over thermal microwave networks~\cite{Xiang2017,Vermersch2017}. As demonstrated in earlier work~\cite{Zhong2019}, the communication architecture here can also use itinerant photons to perform high-fidelity quantum state transfers.

\textbf{Quantum state and process tomography.}
The density matrices of the Bell state and the GHZ states are characterised using quantum state tomography~\cite{Steffen2006}. After the state preparation and transfer, gates from the set $\{I, X/2, Y/2\}$ are applied to each qubit before the simultaneous readout of all qubits; the measured probabilities are corrected for readout errors, and the density matrix is reconstructed numerically. We use CVX, a Matlab package for specifying and solving convex programs, to reconstruct the density matrix while constraining it to be Hermitian, to have unit trace, and to be positive semidefinite. The single-shot simultaneous readout of the qubits is repeated $3\times10^3$ times to obtain the measured probabilities; the state tomography is run repeatedly, in each repeat we reconstruct the density matrix and obtain the state fidelity. The fidelities and uncertainties of the quantum states correspond to the mean and standard deviation of 100 repeated measurements.

Quantum process tomography~\cite{Neeley2008} for the state transfer is carried out by preparing $Q_2^A$ in the input states $\{|g\rangle,(|g\rangle-i|e\rangle)/\sqrt{2}, (|g\rangle+|e\rangle)/\sqrt{2}, |e\rangle\}$, then performing the quantum state transfer process. The corresponding outcome density matrix in $Q_2^B$ is measured using quantum state tomography as described above. The process matrix is reconstructed using the input and outcome density matrices, using the CVX package to constrain it to be Hermitian, unit trace, and positive semidefinite.

\subsection*{Data availability}
The data that support the plots within this paper and other findings of this study are available from the corresponding author upon reasonable request.
\end{document}



\title{Supplementary Information for ``Deterministic multi-qubit entanglement in a quantum network''}
\author{Youpeng Zhong}
\thanks{Present Address: Shenzhen Institute for Quantum Science and Engineering, Southern University of Science and Technology, Shenzhen 518055, China}
\affiliation{Pritzker School of Molecular Engineering, University of Chicago, Chicago IL 60637, USA}
\author{Hung-Shen Chang}
\affiliation{Pritzker School of Molecular Engineering, University of Chicago, Chicago IL 60637, USA}
\author{Audrey Bienfait}
\thanks{Present Address: Universit\'{e} de Lyon, ENS de Lyon, Universit\'{e} Claude Bernard, CNRS, Laboratoire de Physique, F-69342 Lyon, France}
\affiliation{Pritzker School of Molecular Engineering, University of Chicago, Chicago IL 60637, USA}

\author{\'Etienne Dumur}
\thanks{Present Address: Universit\'{e} Grenoble Alpes, CEA, INAC-Pheliqs, 38000 Grenoble, France}
\affiliation{Pritzker School of Molecular Engineering, University of Chicago, Chicago IL 60637, USA}
\affiliation{Center for Molecular Engineering and Material Science Division, Argonne National Laboratory, Argonne IL 60439, USA}
\author{Ming-Han Chou}
\affiliation{Pritzker School of Molecular Engineering, University of Chicago, Chicago IL 60637, USA}
\affiliation{Department of Physics, University of Chicago, Chicago IL 60637, USA}
\author{Christopher R. Conner}
\affiliation{Pritzker School of Molecular Engineering, University of Chicago, Chicago IL 60637, USA}
\author{Joel Grebel}
\affiliation{Pritzker School of Molecular Engineering, University of Chicago, Chicago IL 60637, USA}
\author{Rhys G. Povey}
\affiliation{Pritzker School of Molecular Engineering, University of Chicago, Chicago IL 60637, USA}
\affiliation{Department of Physics, University of Chicago, Chicago IL 60637, USA}
\author{Haoxiong Yan}
\affiliation{Pritzker School of Molecular Engineering, University of Chicago, Chicago IL 60637, USA}
\author{David I. Schuster}
\affiliation{Department of Physics, University of Chicago, Chicago IL 60637, USA}
\affiliation{Pritzker School of Molecular Engineering, University of Chicago, Chicago IL 60637, USA}
\author{Andrew N. Cleland}
\email{Corresponding author; anc@uchicago.edu}
\affiliation{Pritzker School of Molecular Engineering, University of Chicago, Chicago IL 60637, USA}
\affiliation{Center for Molecular Engineering and Material Science Division, Argonne National Laboratory, Argonne IL 60439, USA}

\maketitle

\setcounter{equation}{0}
\setcounter{figure}{0}
\setcounter{table}{0}
\setcounter{page}{1}

\renewcommand{\theequation}{S\arabic{equation}}
\renewcommand{\thefigure}{S\arabic{figure}}
\renewcommand{\thetable}{S\arabic{table}}

\section{Device fabrication}
The device fabrication recipe is adapted from Refs.~\onlinecite{Zhong2019,Kelly2015thesis,Dunsworth2018}, with some modifications to simplify the fabrication of the air-bridge crossovers.

Fabrication steps preceding definition of the qubit and coupler Josephson junctions are done on a 100~mm-diameter sapphire wafer. The wafer is then cut into four quarters, allowing for more attempts for the more delicate junction fabrication.
\begin{enumerate}
\item $100$~nm Al base layer deposition using electron beam evaporation.
\item Base layer photolithography and dry etch with BCl$_3$/Cl$_2$/Ar inductively coupled plasma (ICP). This defines the qubit capacitors, the tunable coupler wiring, and the readout and control circuitry.
\item $200$~nm crossover scaffold SiO$_2$ deposition using photolithography, electron beam evaporation and liftoff.
\item $10$~nm/$150$~nm Ti/Au alignment mark layer deposition using photolithography, electron beam evaporation and liftoff.
\item Josephson junction deposition using the Dolan bridge method \cite{dolan1977} using shadow evaporation and liftoff, using a PMMA/MAA bilayer and electron beam lithography. The Al evaporated in this step does not have any galvanic contact with the base layer wiring.
\item $300$~nm crossover and bandage layer: Al liftoff deposition, preceded by an {\it in situ} Ar ion mill. This step~\cite{Dunsworth2017} creates the top Al layer for crossovers, as well as establishes galvanic connections between the base wiring Al from step 1 and the Josephson junctions defined in step 5.
\item Vapor HF etch to remove the SiO$_2$ scaffold underlying the Al crossovers from step 3.
\end{enumerate}

We use $0.9~\mu$m I-line photoresist AZ MiR 703 for all photolithography steps. The base layer lithography (step 2) uses AZ 300 MIF developer. The other steps (step 3, 4 and 6) use AZ 1:1 developer, which does not attack aluminum. Because the SiO$_2$ scaffold layer (step 3) and the crossover layer (step 6) here involve much thinner deposited layers than those in Refs.~\onlinecite{Zhong2019,Dunsworth2018}, a thick layer of AZ 703 photoresist is sufficient for the lift-off process, which greatly simplifies the fabrication recipe, as compared to the use of tri-layer positive photoresist in Ref.~\onlinecite{Dunsworth2018} or negative photoresist in Ref.~\onlinecite{Zhong2019}. Furthermore, the crossover layer is now merged with the bandage layer~\cite{Dunsworth2017} (step 6) here, further simplifying the fabrication process. Note the air-bridge is mechanically fragile and cannot sustain sonication.

\section{Cable-chip wirebond connections}\label{sec:wirebond}
\begin{figure}[H]
  \centering
  \includegraphics[width=0.8\textwidth]{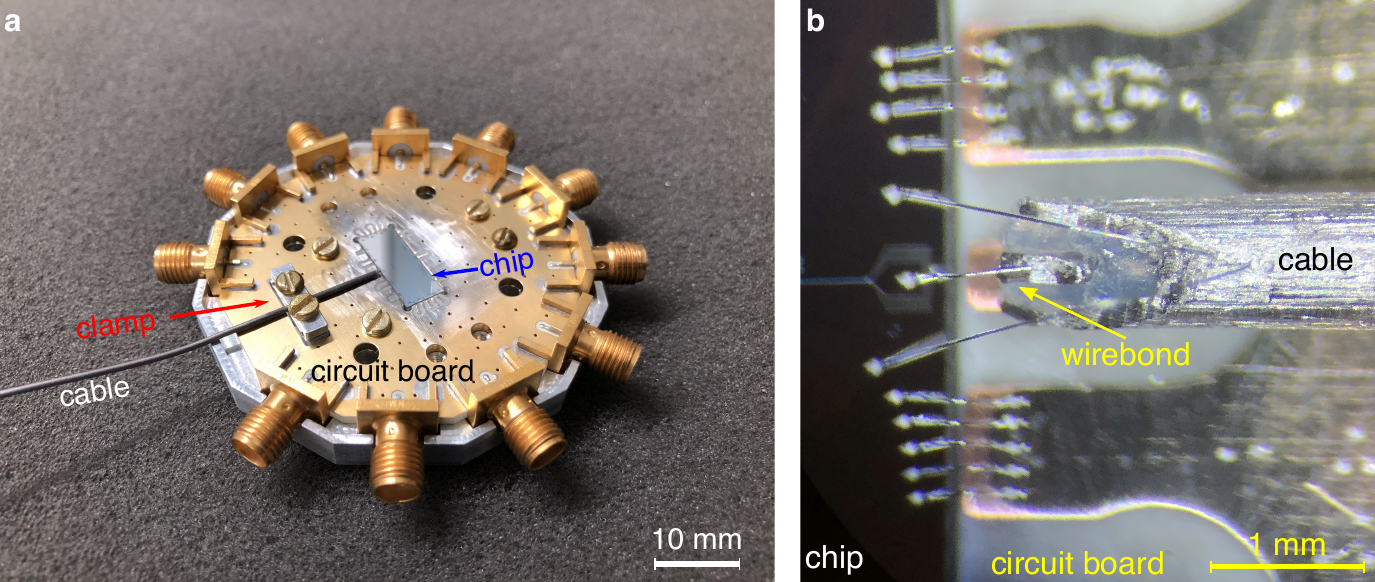}
  \caption{\label{packaging} Cable-chip wirebond connection.
  {\bf a,} Image of the sample holder consisting of a gold-plated printed circuit board, non-magnetic SMA connectors and an aluminum enclosure (the top part of which is removed here). The NbTi cable is held close to the chip, then firmly clamped to the sample holder.
  {\bf b,} Image showing the wirebond connection between the coaxial cable and the processor chip.
  }
\end{figure}

In this experiment, we use a 1 meter long niobium-titanium (NbTi) superconducting coaxial cable (manufacturer: COAX CO., LTD., part number: SC-086/50-NbTi-NbTi) to connect the two superconducting quantum processors. To achieve a high-quality communication channel connection, we avoid the use of normal-metal connectors (e.g. SMA connectors) and instead use $25~\mu$m diameter aluminum wirebonds to connect the cable directly to the superconducting processor chip. A sample holder specifically designed for this purpose is used here, as shown in Fig.~\ref{packaging}a. The NbTi coaxial cable is held close to the processor chip and then firmly clamped on the sample holder with a clamp. The top part of the outer conductor and the PTFE dielectric is removed with a sharp blade to expose the inner conductor at the end of the cable, as shown in Fig.~\ref{packaging}b. The top surfaces of the inner and outer conductors are carefully scraped with a sharp blade to create a flat surface for wirebonding.

\begin{figure}[H]
  \centering
  \includegraphics[width=0.8\textwidth]{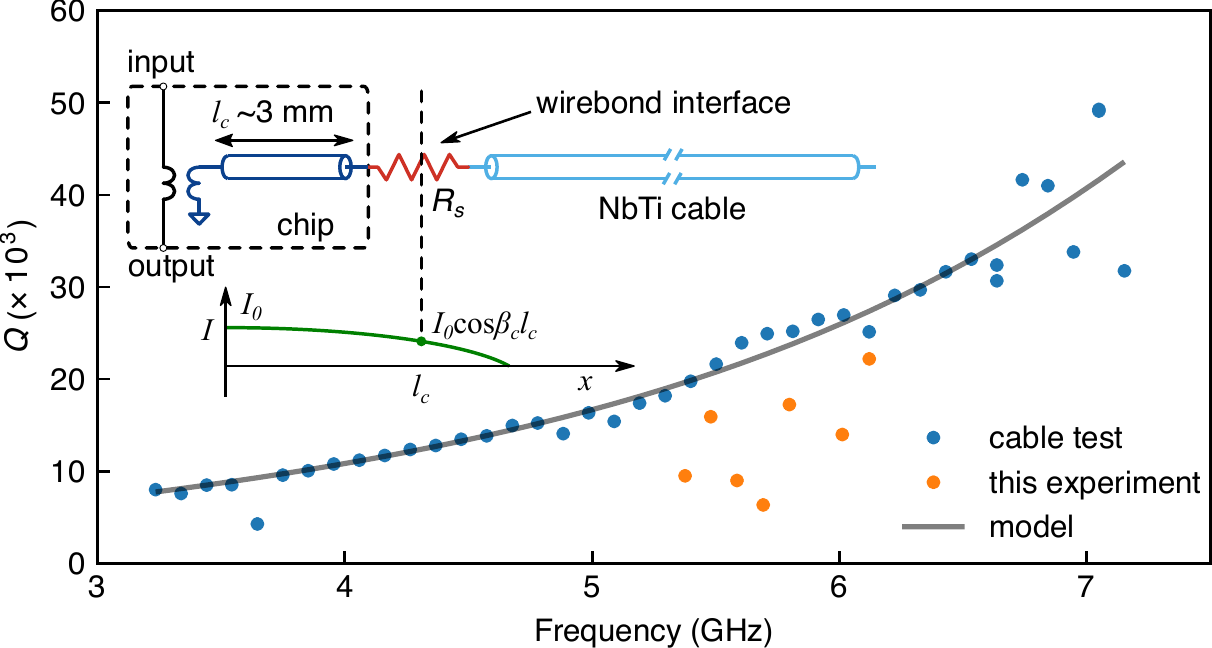}
  \caption{\label{wirebond_loss}
  Channel loss from the wirebond connection.
  Inset: Schematic of independent measurement of coaxial cable loss,
  where the cable is wirebonded to a short CPW line of length $\ell_c\sim 3$~mm on a test chip.
  The loss in the wirebond interface is modeled as a series resistance $R_s$. The current distribution of a standing mode is assumed to follow a simple cosine shape along the CPW line, ignoring the transition in the wirebond interface.
  Blue dots represent the $Q$ of each standing mode measured in this experiment; orange dots are from experiments in main text. Grey line is numerical model with $R_s = 0.38~\Omega$ and $Q_0 = 90.9 \times 10^3$.
}
\end{figure}

The quality factor of the communication channel varies in different assembled devices,  depending strongly on the quality of the wirebond connection. To further explore the loss mechanism in the communication channel, we performed a separate cable test experiment, where we directly wirebond the NbTi cable to a short coplanar waveguide (CPW) line of length $\ell_c \approx 3$~mm on a test chip, see Fig.~\ref{wirebond_loss}. A network analyzer measurement is carried out~\cite{Megrant2012}, yielding the quality factor $Q$ of each standing mode (blue dots). We observe a clear trend of $Q$ increasing with frequency in this cable test. For comparison, we also plot the $Q$ of the standing modes measured in the experiment in the main text (orange dots), and see a similar trend, except some modes have a significantly lower $Q$, likely due to spurious two-level state (TLS) defects near the resonant frequency. The overall frequency dependence is consistent with a resistive dissipation channel $R_s$ in the wirebond interface, as shown inset in Fig.~\ref{wirebond_loss}. This could originate from a thick oxide barrier layer on the NbTi surface. For simplicity, we assume the current of a standing cable mode follows a simple cosine shape along the CPW line. In the wirebond interface, the current is $I_0\cos(\beta_c\ell_c)$ where $I_0$ is the current amplitude at the shorted end, and $\beta_c$ is the propagation constant for the CPW line. This current gives a power loss $P_{\rm loss} = I_0^2 \cos^2(\beta_c\ell_c) R_s$, corresponding to a quality factor of~\cite{Pozar}
\begin{equation}\label{Qloss}
   Q_{\rm loss} = \omega_m \frac{L_m I_0^2}{P_{\rm loss}} = \omega_m \frac{L_m}{\cos^2(\beta_c\ell_c)R_s},
\end{equation}
where $\omega_m/2\pi$ is the standing mode frequency and $L_m$ the lumped element inductance of the mode (see section \ref{sec:mmode}). The $Q$ of the standing mode is then given by
\begin{equation}\label{Qtotal}
    1/Q = 1/ Q_{\rm loss} + 1/Q_0,
\end{equation}
where $Q_0$ is the cable's intrinsic quality factor. Fitting this model with the cable test data, we obtain $R_s=0.38~\Omega$ and $Q_0 = 90.9\times10^3$, shown by the grey line in Fig.~\ref{wirebond_loss}.

In Ref.~\onlinecite{Kurpiers2017}, Kurpiers \emph{et al.} reported an intrinsic $Q$ as high as $92 \times 10^3$ for a NbTi coaxial cable made by Keycom Corp., using capacitive coupling. In Ref.~\onlinecite{Burkhart2020}, a NbTi cable of the same kind used in this experiment is capacitively-coupled to a 3D transmon qubit, where typical $Q$'s of order $50 \times 10^3$ with occasional values as high as $160 \times 10^3$ were observed. The cable intrinsic $Q$ values are quite similar, likely limited by the dielectric loss of PTFE at cryogenic temperatures.

According to the model in Fig.~\ref{wirebond_loss}, if we adjust the coupler circuit such that $\ell_c \sim \lambda/4$, where $\lambda$ is the wavelength of the chip standing mode, then $\cos(\beta_c\ell_c) \approx 0$, minimizing the loss through $R_s$ for frequencies close to the resonant frequency of the $\lambda/4$ transformer. Alternatively, if we use a capacitive tunable coupler design~\cite{Yan2018}, then the channel is open on both ends, and the loss through $R_s$ will be small, as long as $\ell_c \ll \lambda/4$. Another approach is to minimize $R_s$ by using an Al cable instead of NbTi, although cables clad in Al are not easily available.

\section{Experimental setup}
\begin{figure}[H]
  \centering
  \includegraphics[width=0.6\textwidth]{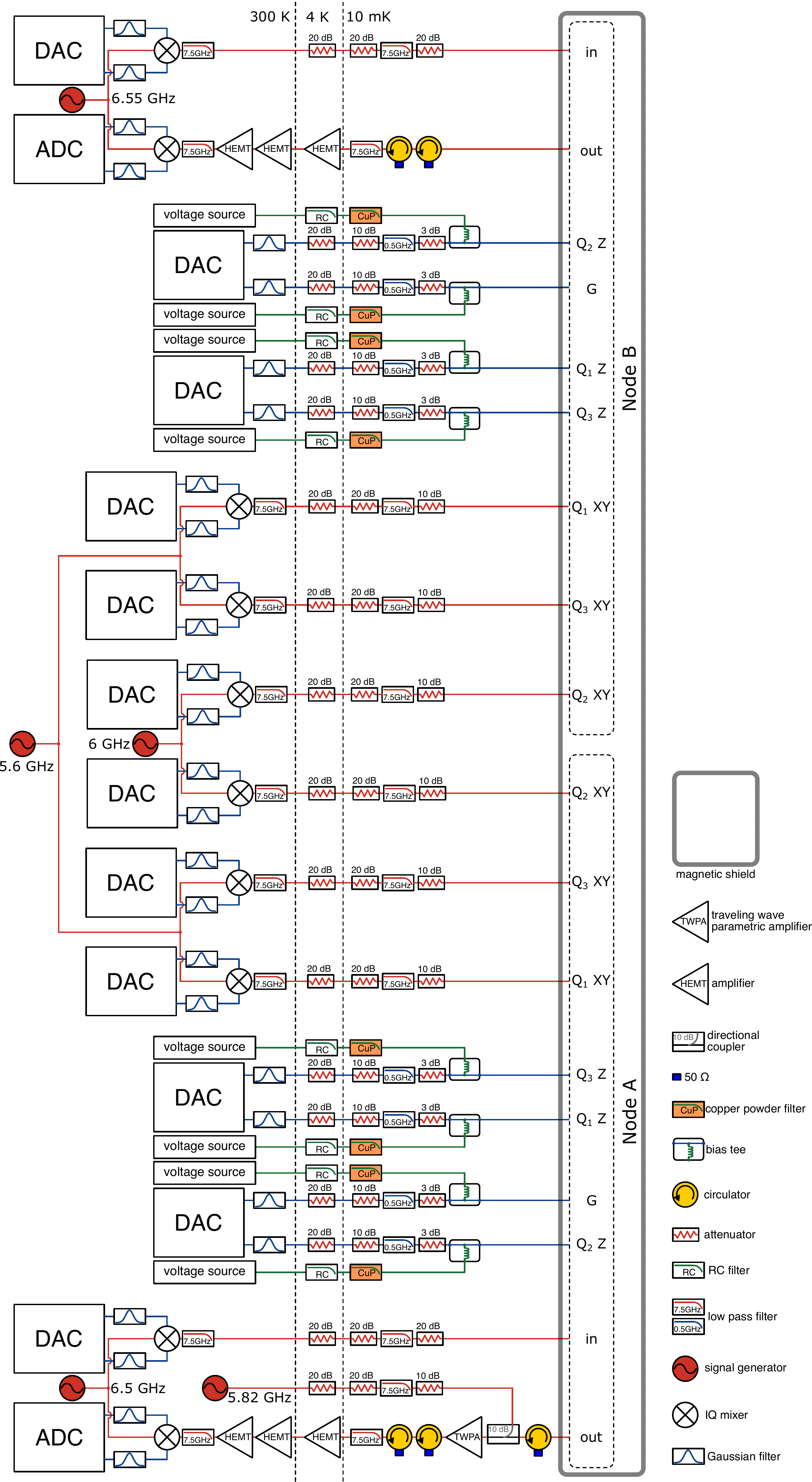}
  \caption{\label{setup}Schematic of the experimental setup.}
\end{figure}

A schematic of the room-temperature electronics and the cryogenic wiring is shown in Fig.~\ref{setup}, similar to that in Ref.~\onlinecite{Zhong2019}. We use custom digital-to-analog converter (DAC) (dual-channel, 14-bit resolution, 1 GS/s sampling rate) and analog-to-digital converter (ADC) (dual-channel, 8-bit resolution, 1 GS/s sampling rate) circuit boards for qubit control and measurement, respectively. Each control signal output and measurement signal input channel is filtered by a custom Gaussian low-pass filter with a $-3$~dB bandwidth of about 250 MHz.

The DAC boards can generate nanosecond-length pulses for fast qubit Z or coupler G control. The fast bias pulse is combined with a direct-current (DC) bias using a bias-tee at the 10 mK stage, where the DC bias line is filtered with an $RC$ filter of $\sim1$~MHz bandwidth at the 4 K stage and a copper powder filter at the 10 mK stage. The DAC dual-channel output can also modulate the envelope of an IQ mixer for qubit XY rotations, or to drive the readout resonator for dispersive measurements. The modulation of the IQ mixer can provide arbitrary waveform output within $\pm 250$~MHz of its local oscillator (LO) frequency. In this experiment, 4 LOs have been used to drive IQ mixers for different purposes, where an LO at 5.6 GHz (6 GHz) carrier frequency is used to control the qubits operating at about 5.5 GHz (5.9 GHz), and an LO at 6.55 GHz (6.5 GHz) carrier frequency is used for the dispersive readout of node $A$ (node $B$) respectively.

The output of the readout microwave signal is first amplified by a traveling wave parametric amplifier (TWPA)~\cite{macklin2015} at the 10 mK stage (node $B$ does not have a TWPA for qubit readout), then amplified by a cryogenic high electron mobility transistor (Low Noise Factory) at the 4 K stage. Two cryogenic circulators with low insertion loss are added between the TWPA and the cryogenic HEMT to block reflections and thermal noise emitted from the input of the cryogenic HEMT. An additional circulator is inserted between the TWPA drive line and the processor, to avoid any unexpected excitation of the qubits from the TWPA drive signal. The cryogenic HEMT output is further amplified by two room-temperature HEMT amplifiers (Miteq Corp.), then down-converted with an IQ mixer and captured by an ADC board.

The ADC board can perform on-board multi-channel demodulation of the captured waveform, yielding a single complex value $\widetilde{I}+i\widetilde{Q}$ in the phase space for each demodulation channel from a single measurement. This allows for the simultaneous readout of multiple qubits using frequency multiplexing~\cite{Chen2012}. With calibrated discrimination criteria in the $\widetilde{I}-\widetilde{Q}$ plane, a $|g\rangle$ or $|e\rangle$ state can be assigned to each $\widetilde{I}+i\widetilde{Q}$ value. Repeating this single-shot measurement several thousand times, we obtain the qubit state probabilities.

\section{Device characterization}
\subsection{Summary of device parameters}
The parameters and typical performance of each qubit are summarized in Table \ref{parameters}. Note the linear inductance from the tunable coupler contributes to the total inductance of the qubit~\cite{Chen2014}, reducing the maximum frequency of $Q_2^n$ by a few hundred MHz. To counteract this effect, the size of $Q_2^n$'s Josephson junctions was increased by 10\% compared to that of the other qubits. The linear inductance from the coupler also weakens the nonlinearity of $Q_2^n$ by about 70 MHz~\cite{Chen2014}, which in turn affects the dispersive shift~\cite{Koch2007}. The readout duration for $Q_2^n$ is correspondingly increased to compensate for this effect.

\begin{table}[H]
\begin{center}
\begin{tabular}{|l |c | c| c |c |c |c |c |c |c |c |c |c}
  \hline
  \hline
   & $f_{eg}^{\rm max}$ (GHz) & $f_{eg}^{\rm idle}$ (GHz) & $\eta$ (GHz)& $T_1$ ($\mu$s) & $T_\phi$ ($\mu$s) & $f_{rr}$ (GHz) & $\tau_{rr}$ (ns) & $F_g$ & $F_e$  \\
  \hline
  $Q_1^A$ & 6.04 & 5.5050 & -0.23 & 12 & 3.4  & 6.5032 & 250 & 0.982 & 0.944\\
  $Q_2^A$ & 6.14 & 5.870 & -0.15 &  7 & 3.8  & 6.5490 & 350 & 0.981 & 0.935\\
  $Q_3^A$ & 6.03 & 5.4882 & -0.23 & 7 &  3.8  & 6.6045 & 300 & 0.985 & 0.942\\
  $Q_1^B$ & 6.08 & 5.4655 & -0.23 & 29 & 4.2  & 6.5065 & 300 & 0.995 & 0.955\\
  $Q_2^B$ & 6.25 & 5.8950 & -0.16 & 11 & 4.4  & 6.5560 & 450 & 0.973 & 0.947\\
  $Q_3^B$ & 6.16 & 5.4835 & -0.23 & 20 & 2.9  & 6.6095 & 300 & 0.984 & 0.953\\
  \hline
  \hline
\end{tabular}
\end{center}
\caption{\label{parameters} Qubit parameters. Here $f_{eg}^{\rm max}$ is the qubit maximum frequency, $f_{eg}^{\rm idle}$ is the qubit idle frequency, $\eta$ is the qubit nonlinearity, $T_1$ and $T_\phi$ are the qubit lifetime and pure dephasing time at the idle frequency respectively, $f_{rr}$ is the readout resonator frequency, $\tau_{rr}$ is the readout length, $F_g$ and $F_e$ are the readout fidelity of the $|g\rangle$ and $|e\rangle$ states respectively.}
\end{table}

To achieve a fast dispersive readout without introducing strong Purcell decay~\cite{Houck2008}, we placed a Purcell filter between the readout resonators and the readout line. The Purcell filter is essentially a shorted half-wavelength coplanar waveguide resonator, similar to that used in Refs.~\onlinecite{Satzinger2018,Bienfait2019,Jeffrey2014}. The filter has a resonant frequency of about 6.5 GHz, a weak coupling to the input port (coupling $Q_c\sim2000$) and a strong coupling to the output port (coupling $Q_c\sim25$). With this element, we are able to perform high-fidelity qubit readout in about 300 ns, even absent a TWPA or parametric amplifier~\cite{Vijay2011}. The readout fidelity for the ground state $|g\rangle$ is $\sim0.98$, primarily limited by the separation error and spurious excitations~\cite{Walter2017}. The readout fidelity of the excited state $|e\rangle$ is $\sim 0.95$, primarily limited by the lifetime of the qubit.

\subsection{Single qubit gate characterization}
\begin{figure}[H]
  \centering
  \includegraphics[width=0.6\textwidth]{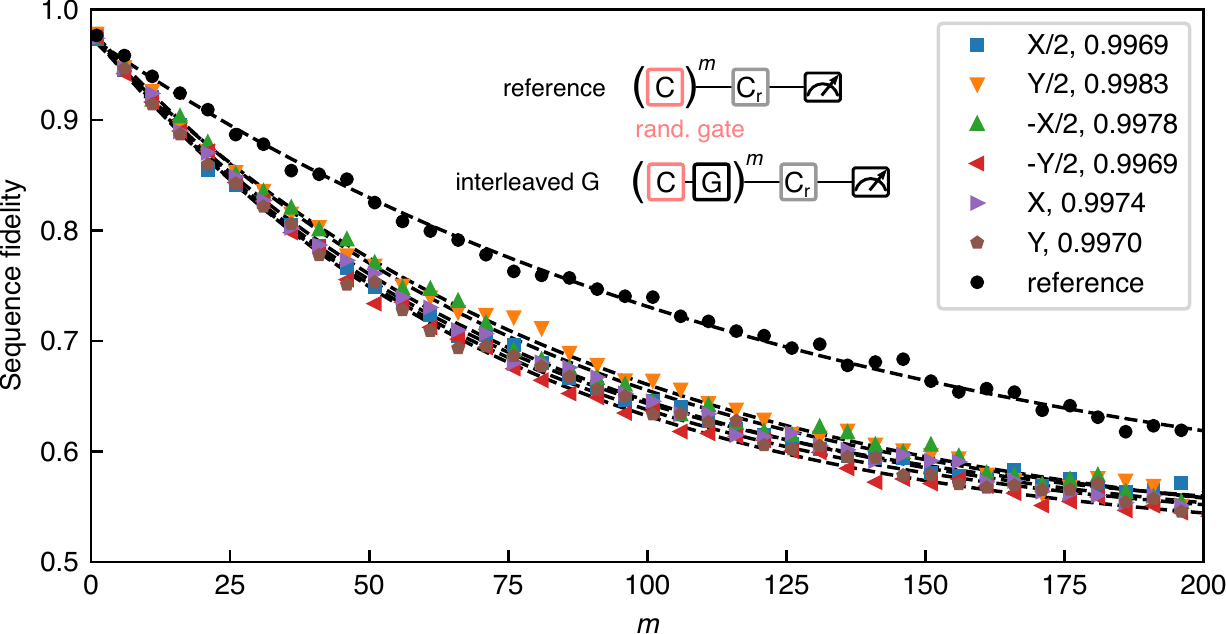}
  \caption{\label{Q1ARB} Single qubit randomized benchmarking for $Q_1^A$. Here $C$ is a random Clifford gate and $C_r$ is a Clifford gate that ideally restores the quantum state after the random gate sequence.}
\end{figure}

We characterize the single qubit gate fidelities using Clifford-based randomized benchmarking (RB)~\cite{Ryan2009,Brown2011,Barends2014}. A typical RB for $Q_1^A$ is shown in Fig.~\ref{Q1ARB}. Table~\ref{sqgates} summarizes the typical single qubit gate fidelities for all qubits in this experiment.

\begin{table}[H]
\begin{center}
\begin{tabular}{|l |c | c| c |c |c |c |c |c |c}
  \hline
  \hline
   & X/2 & -X/2 & Y/2 & -Y/2 & X & Y &average \\
  \hline
  $Q_1^A$ & 0.9969 & 0.9978 & 0.9983 & 0.9969 & 0.9974  & 0.9970 & 0.9974 \\
  $Q_2^A$ & 0.9979 & 0.9969 & 0.9971 & 0.9980 & 0.9976  & 0.9973 & 0.9975 \\
  $Q_3^A$ & 0.9987 & 0.9985 & 0.9988 & 0.9970 & 0.9953  & 0.9983 & 0.9978\\
  $Q_1^B$ & 0.9973 & 0.9990 & 0.9978 & 0.9976 & 0.9985  & 0.9981 & 0.9981\\
  $Q_2^B$ & 0.9982 & 0.9951 & 0.9965 & 0.9959 & 0.9937  & 0.9969 & 0.9961\\
  $Q_3^B$ & 0.9947 & 0.9968 & 0.9995 & 0.9967 & 0.9932  & 0.9983 & 0.9965\\
  \hline
  \hline
\end{tabular}
\end{center}
\caption{\label{sqgates} Single qubit gate fidelities for all qubits in this experiment, as determined by randomized benchmarking.}
\end{table}

\subsection{iSWAP and CZ gates}\label{sec:cz}
\begin{figure}[H]
  \centering
  \includegraphics[width=0.8\textwidth]{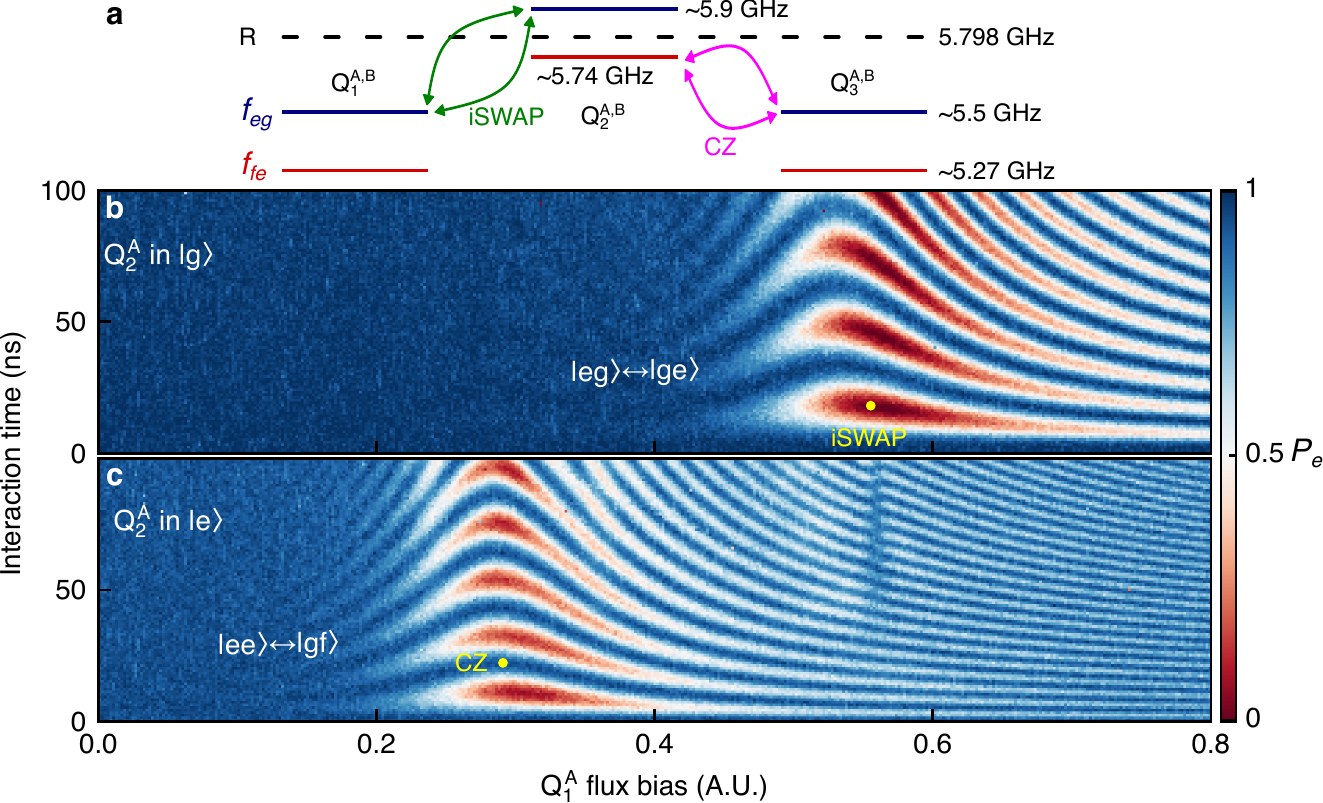}
  \caption{\label{cz}Implementation of two-qubit gates.
  {\bf a,} Transition frequency diagram of the three qubits in each node, showing both the $g-e$ ($f_{ge}$) and $e-f$ ($f_{ef}$) transitions. The center qubits $Q_2^n$ ($n=A,B$) are operated at $\sim5.9$~GHz, and the side qubits $Q_{1,3}^n$ are operated at $\sim5.5$~GHz.
  {\bf b,} Vacuum Rabi oscillations for the two-qubit $|eg\rangle-|ge\rangle$ transition between $Q_1^A$ and $Q_2^A$. An iSWAP gate can be implemented by enabling this oscillation for a duration of 15~ns, as marked by the yellow dot.
  {\bf c,} Vacuum Rabi oscillations for the $|ee\rangle-|gf\rangle$ transition between $Q_1^A$ and $Q_2^A$. A CZ gate can be implemented by enabling this oscillation for a duration of 21~ns, as marked by the yellow dot.
  }
\end{figure}

The transition frequency diagram of the three qubits in each node is shown in Fig.~\ref{cz}a. The central qubits $Q_2^n$ ($n=A,B$) operate with their $g-e$ transition $f_{eg}$ (blue) at $\sim 5.9$~GHz, while the other two qubits $Q_{1,3}^n$ operate at $f_{eg} \sim 5.5$~GHz (slightly detuned from one another). The $e-f$ transition $f_{ef}$ (red) is around 5.9 GHz for $Q_2^n$ and around 5.5 GHz for $Q_{1,3}^n$. With $\sim 0.4$~GHz detuning ($\sim 0.24$~GHz detuning between the $|ee\rangle$-$|gf\rangle$ transition), the residual coupling between adjacent qubits is very small. During the quantum state transfer operation, $Q_2^n$ is tuned to 5.798 GHz to resonantly interact with the communication mode $R$. At this frequency, the detuning between $Q_2^n$ and the side qubits $Q_{1,3}^n$ is not small enough to avoid unwanted stray coupling, so we apply detuning pulses to $Q_{1,3}^n$ to reduce their transition frequencies by about 200 MHz during this process.

To swap a quantum state from $Q_{1,3}^n$ to $Q_2^n$ (initially in the $|g\rangle$ state), we bias $Q_{1,3}^n$ so its $g-e$ transition is resonant with that of $Q_2^n$, initiating vacuum Rabi oscillation between the $|eg\rangle$ and $|ge\rangle$ states, as shown in Fig.~\ref{cz}b. At $\tau_{\rm swap} = \pi/2 g^n_{j,2} = 15$~ns, we complete the $|eg\rangle \rightarrow -i|ge\rangle$ iSWAP process. Ideally, the $|ee\rangle$ state is unchanged under this gate, but as shown in Fig.~\ref{cz}c, due to the weak nonlinearity of $Q_2^n$, if both qubits are in the $|e\rangle$ state, stray coupling between the $|gf\rangle$ state and the $|ee\rangle$ state can cause state leakage during the iSWAP gate. Fortunately, in this experiment, the receiver qubit is ideally always in its $|g\rangle$ state when we transfer states using the iSWAP gate, so this state leakage is not a concern.

\begin{figure}[H]
  \centering
  \includegraphics[width=0.8\textwidth]{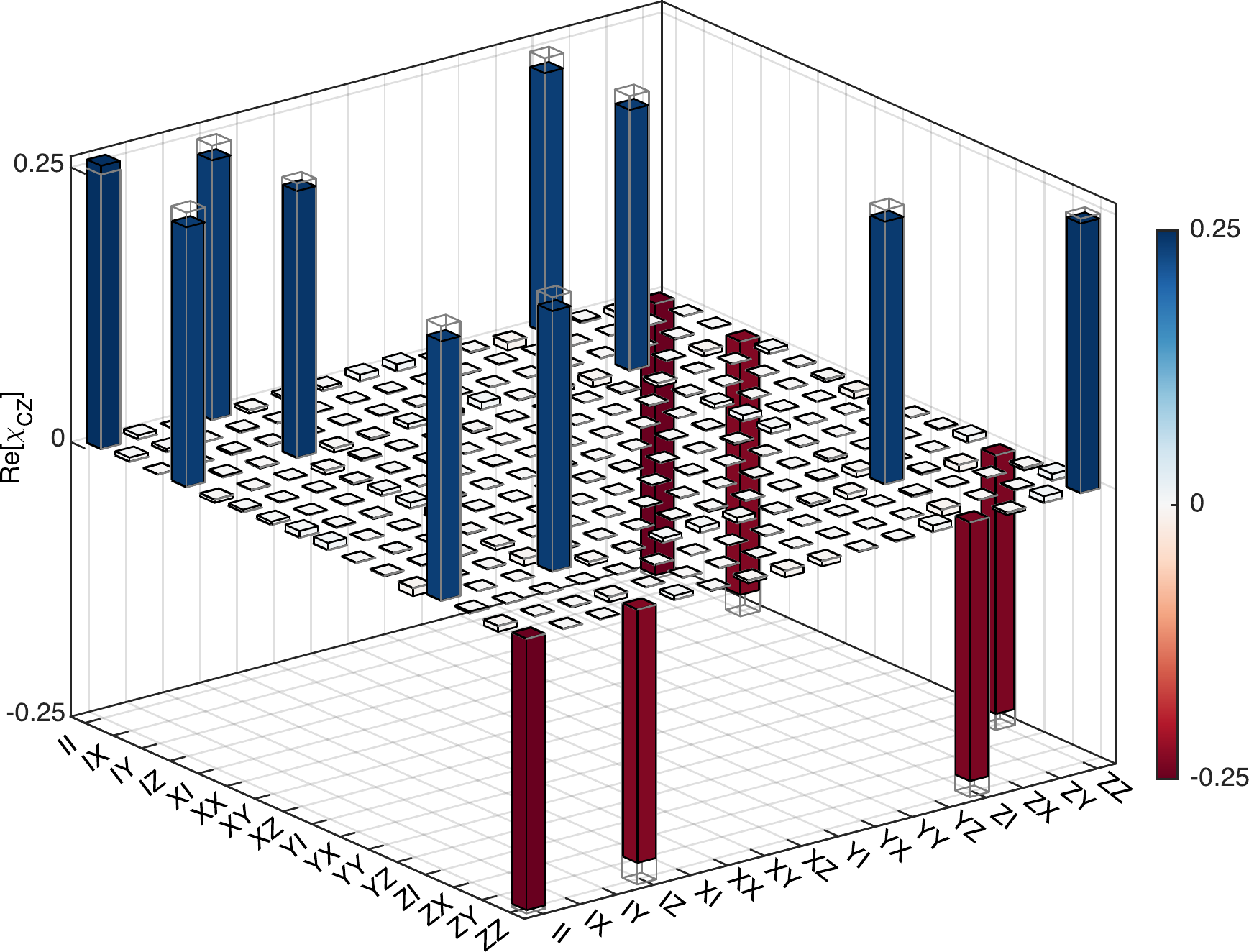}
  \caption{\label{chi_cz}Quantum process tomography of the CZ gate between $Q_1^A$ and $Q_2^A$. The solid color and gray outline bars are for the measured and ideal values respectively. The CZ gate has a process fidelity of $0.958\pm 0.007$.}
\end{figure}

To characterize the transfer efficiency of the iSWAP gate, we compare the $Q_2^n$ $|e\rangle$ final state probability $P_e$ from two experiments: In one experiment, we apply a $\pi$ pulse to $Q_2^n$ directly, and then measure; in the other experiment, we apply a $\pi$ pulse to $Q_{1,3}^n$ and then transfer the excitation to $Q_2^n$ using an iSWAP gate, followed by measurement. These two experiments are carried out back-to-back and repeated 1000 times. We compare the average $\langle P_e \rangle$ from the two experiments, and find that the iSWAP gate has a transfer efficiency $\eta_{\rm iSWAP}\approx 0.99$.

\begin{table}[H]
\begin{center}
\begin{tabular}{|c |c | c| c | c | c|}
  \hline
  \hline
   &$Q_1^A-Q_2^A$ & $Q_3^A-Q_2^A$ & $Q_1^B-Q_2^B$ & $Q_3^B-Q_2^B$ & average\\
  \hline
  $\mathcal{F}_{CZ}$ & 0.958(7) & 0.945(8) & 0.952(5) & 0.944(7) & 0.950(6)\\
  \hline
  \hline
\end{tabular}
\end{center}
\caption{\label{czgates} CZ gate fidelities, determined by process tomography.}
\end{table}

The CZ gate here is implemented utilizing the $|f\rangle$ state of $Q_2^n$, as proposed in Ref.~\onlinecite{Strauch2003} and demonstrated in Refs.~\onlinecite{Dicarlo2010,Yamamoto2010}. When biasing $Q_{1,3}^n$ to be resonant with the $|e\rangle$-$|f\rangle$ transition frequency of $Q_2^n$, a vacuum Rabi oscillation between the $|ee\rangle$ and $|gf\rangle$ state can be observed, as shown in Fig.~\ref{cz}c. If the interaction is turned on for $\tau_{\rm CZ}=\pi/\sqrt{2}g^n_{j,2} \approx 21$~ns, $j=1,3$, the quantum state completes an $|ee\rangle\rightarrow-i|gf\rangle\rightarrow-|ee\rangle$ round trip and acquires a $\pi$ phase relative to the other states, as required for this gate~\cite{Strauch2003}.

We perform quantum process tomography to characterize the CZ gate between $Q_1^A$ and $Q_2^A$ here, yielding the process matrix $\chi_{CZ}$ shown in Fig.~\ref{chi_cz}, with a process fidelity of $\mathcal{F}_{CZ} = {\rm Tr}(\chi_{CZ}\cdot\chi_{CZ,\rm ideal})=0.958\pm0.007$, here $\chi_{CZ,\rm ideal}$ is the process matrix for the ideal CZ gate, and the error bar is the standard deviation of repeated measurements. The fidelities of all the CZ gates are summarized in Table~\ref{czgates}, with an average fidelity of $0.950\pm 0.006$, here the error bar is the standard deviation of the four CZ gate fidelities.

\begin{figure}[H]
  \centering
  \includegraphics[width=0.8\textwidth]{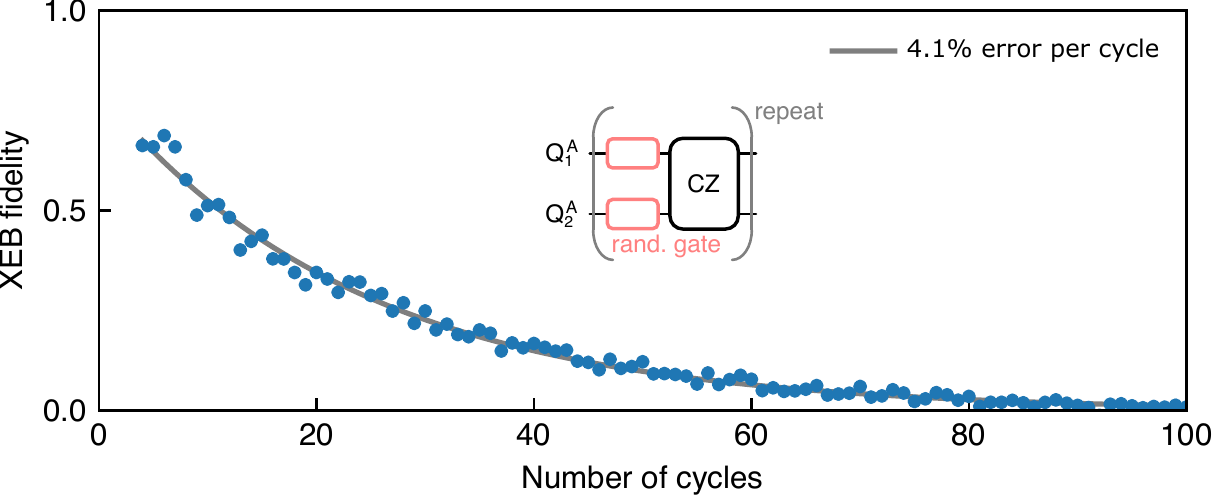}
  \caption{\label{CZ_XEB}Cross-entropy benchmarking of the CZ gate, with an error of 4.1\% per cycle.}
\end{figure}

In addition to quantum process tomography, we use the cross-entropy benchmarking (XEB) technique~\cite{Arute2019} to estimate the fidelity of the CZ gate, where we measure an error of 4.1\% per cycle. Subtracting the single-qubit gate errors, the CZ gate fidelity is 0.964, in good agreement with the process tomography fidelity. Note that CZ gate fidelities $>0.99$ can be achieved with an optimized adiabatic gate~\cite{Barends2014,Martinis2014} or tunable coupling~\cite{Chen2014,Arute2019,Xu2020}.

A dynamic phase is accumulated by each qubit when performing the two-qubit gates, due to the change of the qubit frequency during the interaction. This dynamic phase can be physically corrected by applying a calibrated Z rotation. Alternatively, to simplify the control sequence, here we adjust the phase of the tomography pulses to correct for the dynamic phase shift when performing quantum state tomography. Similarly, we adjust the phase of the second $Y/2$ gate on the target qubit to correct for the dynamic phase shift when performing a CNOT gate.

\subsection{Flux crosstalk}
\begin{figure}[H]
  \centering
  \includegraphics[width=0.8\textwidth]{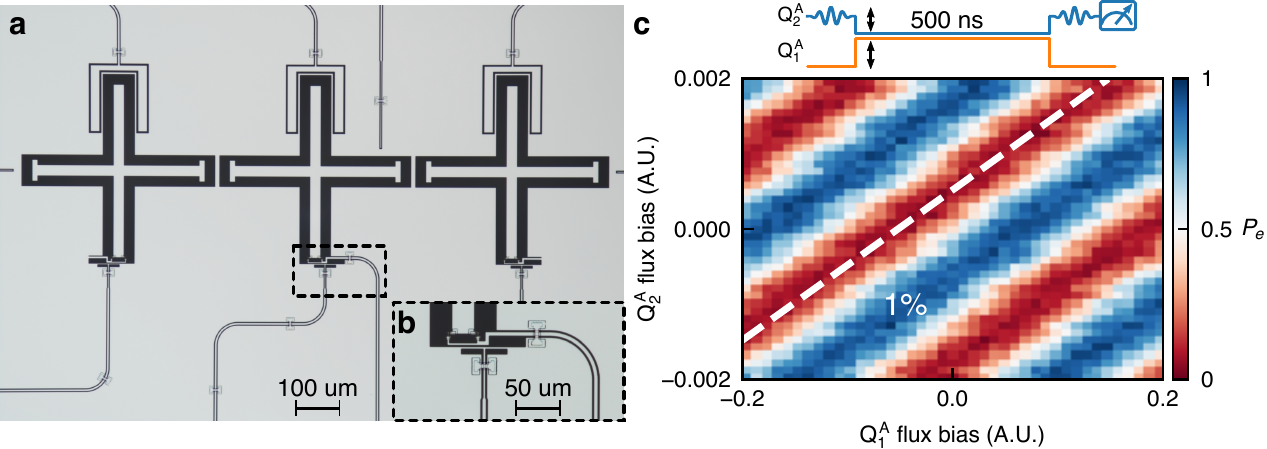}
  \caption{\label{crosstalk}Measurement of qubit and coupler magnetic flux crosstalk.
  {\bf a,} Xmon qubit design adopted from Ref.~\onlinecite{Barends2013}, with a gradiometer flux control line design.
  {\bf b,} Magnified view of the flux control line design.
  {\bf c,} Characterization of the flux crosstalk between $Q_1^A$ and $Q_2^A$, using Ramsey interference.
  The control pulse sequence is shown on top, where the black double-headed arrows represent the effect of the control fluxes varied in the measurement. The white dashed line is a representative contour line of the data, along which the fluxes from the control lines for $Q_1^A$ and $Q_2^A$ cancel one another. The slope of this line, which is 1\% here, represents the flux crosstalk between these two qubits.}
\end{figure}

There are 6 qubits and 2 tunable couplers in the quantum network measured here, each of which has an independent flux control line. It is very important to mitigate the flux crosstalk between these channels, to achieve the highest fidelity qubit control and state transfer. First, we implemented a gradiometer design for the flux control line of each qubit, in order to minimize the flux crosstalk, as shown in Fig.~\ref{crosstalk}a and b. We then measured the flux crosstalk between qubit pairs using Ramsey interference. We find that the crosstalk between neighbouring qubits is about 1\%, where a representative measurement in shown in Fig.~\ref{crosstalk}c. The flux crosstalk between the cable-coupled qubits $Q_2^n$ and their adjustable couplers $G^n$ in each node is estimated to be 3-6\%, using spectroscopy measurements (not shown).

\section{Qubit-cable coupling}\label{sec:mmode}
The $\ell_{cb} = 1$~m long NbTi cable has a specific capacitance $\mathscr{C}_{cb} = 96.2$~pF/m and a specific inductance $\mathscr{L}_{cb} = 240.5$~nH/m (as provided by the cable manufacturer). The cable is galvanically connected to the tunable couplers by a short segment of CPW line of length $\ell_c\approx2$~mm patterned on each quantum processor die. The CPW line has a specific capacitance $\mathscr{C}_{cpw} = 173$~pF/m and specific inductance $\mathscr{L}_{cpw} = 402$~nH/m. The $m^{\rm th}$ standing mode in the CPW-cable-CPW channel can be modeled as a lumped element series $LC$ resonator~\cite{Pozar}, with parameters given by
\begin{eqnarray}
  L_m &\approx& \frac{1}{2} (\mathscr{L}_{cp} \ell_{cp}+2 \mathscr{L}_{cpw} \ell_{c})={\rm121\: nH}, \\
  \omega_m &\approx& m \omega_{\rm FSR},\\
  C_m &=& \frac{1}{\omega_m^2L_m}.
\end{eqnarray}

Each qubit $Q_2^n$ ($n=A,B$) is coupled to the channel via a tunable coupler $G^n$ with the same design as in Ref.~\onlinecite{Zhong2019}. This configuration is accurately modeled~\cite{Chen2014,Geller2015} as a tunable inductance given by
\begin{equation}\label{M}
  M^n_c = \frac{L_g^2}{2L_g+L_w+L_T^n/\cos\delta^n},
\end{equation}
where $\delta^n$ is the phase across the coupler Josephson junction, $L_T^n$ is the coupler junction inductance at $\delta^n=0$, $L_g=0.2$~nH, and $L_w \approx 0.1$ nH represents the stray wiring inductance, which cannot be ignored when $L_T^n$ becomes very small~\cite{Zhong2019}.

In the harmonic limit and assuming weak coupling, the coupling between qubit $Q_2^n$ and the $m^{\rm th}$ mode is~\cite{Chen2014,Geller2015}
\begin{equation}\label{coupling_m}
  g_m^n = -\frac{M^n_c}{2} \, \sqrt{\frac{\omega_m \omega_2^n}{(L_g+L_q^n)(L_g+L_m)}},
\end{equation}
where $L_q^n \approx 8.4$~nH is the qubit $Q_2^n$ inductance and $\omega_2^n/2\pi$ is $Q_2^n$'s operating frequency.
We see that $g_m^n \propto \sqrt{\omega_m} \propto \sqrt{m}$, a well-known result for multi-mode coupling~\cite{Sundaresan2015}. It is experimentally more practical to approximate the coupling by a single value $g^n$, because as the mode number $m \sim 55 \gg 1$ near 5.8~GHz, the variation in $g_m^n$ with $m$ within the frequency range of interest is small.

\begin{figure}[H]
  \begin{center}
  \includegraphics[width=0.8\textwidth]{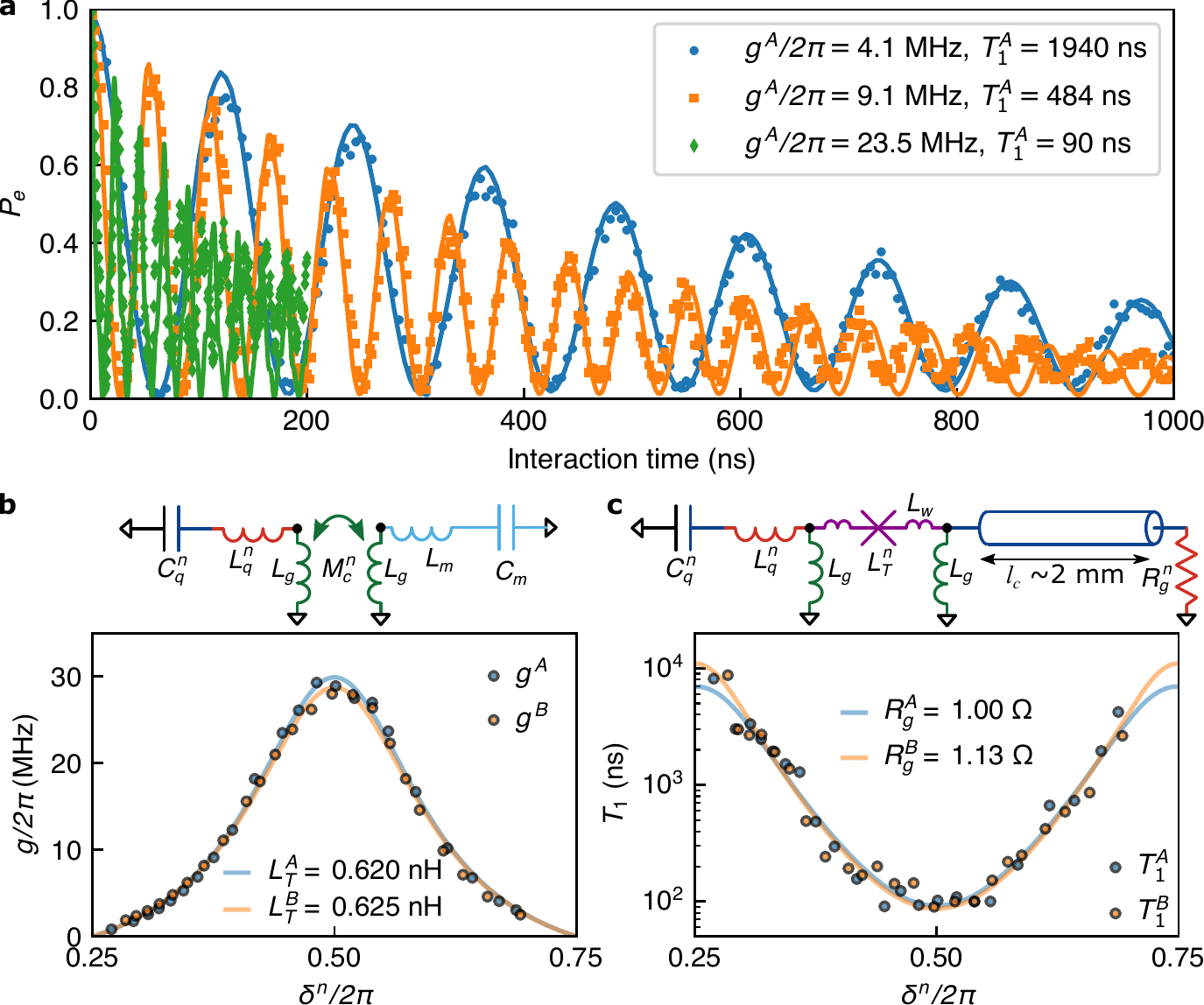}
  \caption{\label{coupling} Tunable coupler characterization.
  {\bf a,} Vacuum Rabi oscillations between qubit $Q_2^A$ and the communication mode $R$ at different coupling strength $g^A/2\pi$. Fitting the data gives $g^A/2\pi$ and the qubit lifetime $T_1^A$ during the interaction.
  {\bf b,} $g^n/2\pi$ versus $\delta^n$. Top: lumped-element linear circuit model for the inductive coupling between the qubit $Q_2^n$ and the communication mode $R$, which is modeled as a series $LC$ resonator.
  {\bf c,} $Q_2^n$ lifetime $T_{1}^n$ versus $\delta^n$ during the interaction. Top: phenomenological circuit model for calculating the qubit loss, assuming a loss channel $R_g^n$ shunting the wirebond connection to ground.}
  \end{center}
\end{figure}

To characterize the tunable couplers, we vary the coupler junction phase $\delta^n$ and tune the qubit $Q_2^n$ to resonantly interact with the communication mode $R$, as shown in Fig.~\ref{coupling}a. The coupling strength $g^n/2\pi$ versus $\delta^n$ is shown in Fig.~\ref{coupling}b, which is obtained by fitting a series of vacuum Rabi oscillations similar to Fig.~\ref{coupling}a (for details of the fitting, see Section~\ref{sec:simu}). We fit the analytical model, Eq.~(\ref{coupling_m}), to the data in Fig.~\ref{coupling}b, and find that $L_T^A = 0.620$~nH and $L_T^B = 0.625$~nH. Maximum coupling occurs at the junction phase $\delta^n = \pi$, where $g^A_{\rm max}/2\pi \approx 29$~MHz and $g^B_{\rm max}/2\pi \approx 28$~MHz. The coupling can be turned off by setting $\delta^n=\pi/2$, making $L_T^n/\cos \delta^n$ very large.

It can be seen from Fig.~\ref{coupling}a that the envelope of the vacuum Rabi oscillation decays faster as the coupling strength increases. This is attributed to the lossy wirebond interface, which not only introduces dissipation to the communication channel, but also affects the qubit coherence. Here we use a phenomenological model, shown at the top of Fig.~\ref{coupling}c, to characterize the qubit loss at different coupling strengths, where we simply assume a lumped resistor $R_g^n$ shunting the wirebond interface to ground. We fit the model with the measured qubit $T_1^n$ (log scale), and find that the model agrees very well with the data, with $R_g^A=1.00$~$\Omega$ and $R_g^B=1.13$~$\Omega$.

\section{Numerical Simulations}\label{sec:simu}
The full quantum system can be modeled with the following rotating-frame, multi-qubit, multi-mode communication channel Hamiltonian:
\begin{eqnarray}\label{H}
  H/\hbar & = & \sum_{i=1,2,3}^{n=A,B} \Delta\omega_{i}^n {\sigma_{i}^n}^\dag \sigma_{i}^n + \sum_{m=1}^{M} \left (m-\frac{M+1}{2} \right ) \omega_{\rm{FSR}} a_m^\dag a_m \\
  &&+\sum_{n=A,B} \sum_{j=1,3} g_{j,2}^n \left (\sigma_{2}^n {\sigma_{j}^n}^\dag + {\sigma_{2}^n}^\dag \sigma_{j}^n \right ) \nonumber\\
  &&+\sum_{m=1}^{M} g^A \left (\sigma_{2}^A a_m^\dag + {\sigma_{2}^A}^\dag a_m \right ) +\sum_{m=1}^{M} (-1)^m g^B \left (\sigma_{2}^B a_m^\dag + {\sigma_{2}^B}^\dag a_m \right)\nonumber,
\end{eqnarray}
where $\sigma_i^n$ and $a_m$ are the annihilation operators for qubit $Q_i^n$ and the $m^{\rm th}$ standing-wave mode respectively, $\Delta\omega_i^n$ is the qubit frequency detuning with respect to the rotating frame frequency, and $M$ is the number of standing modes included in the simulation (always chosen to be an odd number). The rotating frame frequency is set at the center of the standing-mode frequencies, i.e. for mode number $m = (M+1)/2$. Note the sign of $g^B$ alternates with the mode number $m$ due to the parity dependence of the standing wave mode~\cite{Pellizzari1997,Vogell2017}. In this experiment, not all components are involved simultaneously, and in certain cases the full Hamiltonian can be simplified.

In Fig.~2b in the main text, where only $Q_2^A$ and the standing modes are interacting, the Hamiltonian can be simplified to
\begin{eqnarray}\label{H2b}
  H/\hbar & = & \sum_{m=1}^{M} \left (m-\frac{M+1}{2} \right ) \omega_{\rm{FSR}} a_m^\dag a_m+\sum_{m=1}^{M} g^A \left (\sigma_{2}^A a_m^\dag + {\sigma_{2}^A}^\dag a_m \right ),
\end{eqnarray}
where we choose $M=5$ standing modes, with the third mode $m=3$ the communication mode $R$, and $Q_2^A$  is assumed to be on resonant with $R$ such that $\Delta\omega_2^A=0$. Decoherence is taken into account using the Lindblad master equation. The quantum state evolution is calculated using QuTiP~\cite{Johansson2012}. The five standing modes included in the model here have measured lifetimes of 256~ns, 177~ns, 473~ns, 200~ns, and 370~ns respectively. We first compare the numerical simulations using the qubit intrinsic lifetime $T_1=7$~us, and find discrepancies with the data (see the grey line in Fig.~2b of the main text). As discussed in Section~\ref{sec:wirebond}, the loss in the channel is dominated by the wirebond interface. Changing the coupler inductance does not change the participation of the lossy wirebond interface in the channel, so the lifetime of the standing modes should not be affected by the coupling strength. On the other hand, when the coupling is turned on, the qubit is exposed to the lossy wirebond interface, introducing a new loss channel to the qubit coherence. This unwanted side-effect is characterized by the phenomenological circuit model shown in Section~\ref{sec:mmode}. We fit the master equation simulation to the experimental data and find that the qubit $T_1$ is decreased to 1.4~$\mu$s during the interaction (red line in Fig.~2b in the main text). Similarly, we fit a series of vacuum Rabi oscillations, as shown in Fig.~\ref{coupling}a, to obtain the coupling strength $g^n/2\pi$ (Fig.~\ref{coupling}b) and the qubit lifetime $T_1^n$ (Fig.~\ref{coupling}c) at different coupler junction phases $\delta^n$.

In Fig.~2c in the main text, where the side qubits $Q_{1,3}^n$ are tuned far in frequency from $Q_2^n$, the state transfer process can be modeled with the simplified Hamiltonian:
\begin{eqnarray}\label{H2c}
  H/\hbar & = & \sum_{n=A,B} \Delta\omega_{2}^n {\sigma_{2}^n}^\dag \sigma_{2}^n + \sum_{m=1}^{M} \left (m-\frac{M+1}{2} \right ) \omega_{\rm{FSR}} a_m^\dag a_m \\
  &&+\sum_{m=1}^{M} g^A \left (\sigma_{2}^A a_m^\dag + {\sigma_{2}^A}^\dag a_m \right ) +\sum_{m=1}^{M} (-1)^m g^B \left (\sigma_{2}^B a_m^\dag + {\sigma_{2}^B}^\dag a_m \right)\nonumber,
\end{eqnarray}
where we include $M=5$ standing modes in the simulations, and $R$ is the third mode, $m=3$. Ideally, for the hybrid state transfer scheme~\cite{Wang2012}, both qubits $Q_2^n$ should be resonant with $R$, such that $\Delta\omega_2^n=0$, and the coupling $g^A$ and $g^B$ should be set to the same coupling strength $g_0$ simultaneously for a duration $\tau$. In the experiment, we vary the qubit frequencies as well as the relative amplitude and delay between $g_A$ and $g_B$, to optimize the transfer fidelity. It is found that a higher fidelity is achieved with a delay of $\Delta\tau=13$ ns between the initial turn-on for $g_A$ and $g_B$ (in other words, both $g_A$ and $g_B$ are turned on for a duration of $\tau$, but $g^B$ is turned on 13 ns later than $g^A$). With this experimentally-optimized $\Delta\tau$, we fit the model to the data shown in Fig.~2c in the main text, and find that $\Delta \omega_{2}^A/2\pi = -0.95$~MHz, $\Delta \omega_{2}^B/2\pi = -1.79$~MHz, $g_A/2\pi = 4.08$~MHz and $g_B/2\pi = 4.06$~MHz (these are the parameters for the grey line in Fig.~2c in the main text).

In Fig.~3b of the main text, the numerical $\rho^A$ is calculated using the CZ gate process matrix $\chi_{CZ}$ measured in Section~\ref{sec:cz}, assuming the single-qubit rotation gates are ideal (using their measured fidelities has almost no impact on the results). The numerical GHZ state fidelity is 0.938, agreeing well with the experiment. The prepared GHZ state fidelity is primarily limited by the CZ gate fidelity, which could be improved by using an optimized adiabatic gate~\cite{Barends2014,Martinis2014} or using tunable coupling~\cite{Chen2014,Arute2019,Xu2020}. Some one-step GHZ state preparation methods utilizing a common bus resonator may also be able to prepare high-fidelity GHZ states~\cite{Zheng2001,Song2017}. In Fig.~3c of the main text, the numerical $\rho^B$ is calculated by applying the state transfer process $\chi^{\otimes3}$ and the decoherence process to $\rho^A$ from Fig.~3b. The fidelity of $\rho^B$ is primarily limited by the state transfer fidelity $\mathcal{F}^p$, which might be improved by optimizing the coupler circuit design or using a coaxial cable made with different superconducting material, e.g. aluminium, as discussed in Section~\ref{sec:wirebond}.

\begin{figure}[H]
  \centering
  \includegraphics[width=0.9\textwidth]{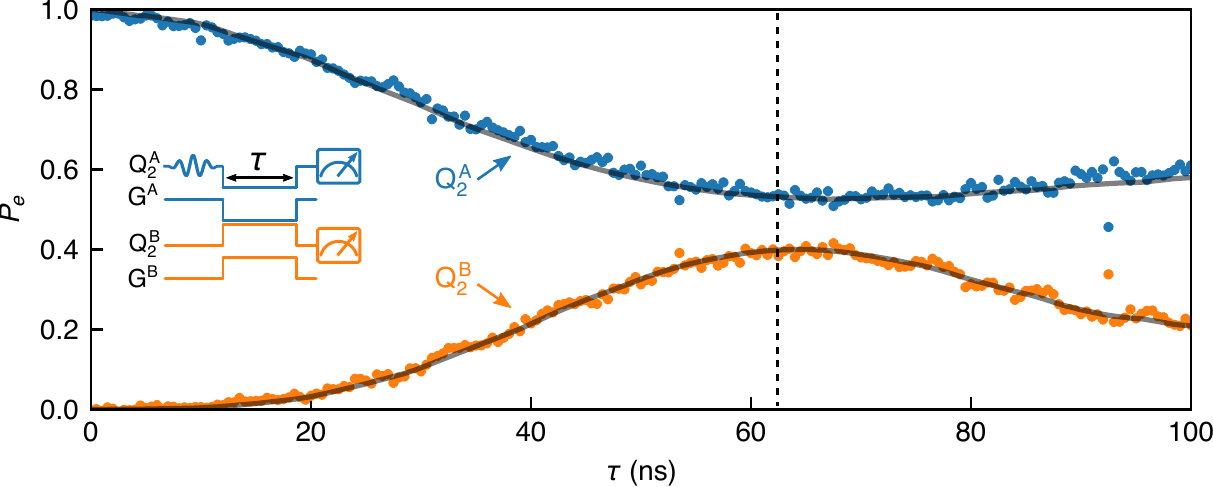}\\
  \caption{The ``ST/2" process, where half a photon is sent from node $A$ to node $B$. The control pulse sequence here is similar to that in Fig.~2c in the main text, except the coupling strength $g^A$ and $g^B$, the interaction time $\tau$ and the delay $\Delta\tau$, are experimentally tuned to optimize the Bell state fidelity. The dashed line marks the point where the Bell state fidelity is optimal.}\label{bellGen}
\end{figure}

In Fig.~4b in the main text, the control pulse for ``ST/2" is similar to that for ``ST" as shown in Fig.~2c inset, except the coupling strength $g^A$ and $g^B$, the interaction time $\tau$ and the delay $\Delta\tau$, are experimentally tuned to optimize the Bell state fidelity. With $\Delta\tau=5$~ns, as determined experimentally, we fit the data in Fig.~\ref{bellGen}, which is similar to Fig.2c in the main text, and obtain $\Delta\omega_{2}^A/2\pi = 4.7$~MHz, $\Delta\omega_{2}^B/2\pi = 5.4$~MHz, $g_A/2\pi = 2.89$~MHz and $g_B/2\pi = 6.11$~MHz. The numerical Bell state fidelity is 0.915, agreeing well with the experiment. This Bell state fidelity is primarily limited by the channel loss.

In Fig.~4c, we calculate the theoretical $\rho_{II}$ by applying $\chi_{CZ}$ (measured in Section~\ref{sec:cz}) to $\rho_I$ from Fig.~4b, assuming the single qubit rotation gates are ideal.

In Fig.~4d, we calculate the theoretical $\rho_{III}$ by applying $\chi_{CZ}$ and decoherence process to $\rho_{II}$ from Fig.~4c, again assuming the single-qubit rotation gates are ideal. The decoherence process is applied to $Q_1^n$ to account for the idling of 70~ns during the application of CNOT gates to $Q_3^n$. The fidelity of $\rho_{II}$ and $\rho_{III}$ is primarily limited by the fidelity of $\rho_I$ and the CZ gates.

\clearpage

\bibliographystyle{naturemag}
\bibliography{bibliography}